# An Ensemble 1D-CNN-LSTM-GRU Model with Data Augmentation for Speech Emotion Recognition

Md. Rayhan Ahmed, Salekul Islam, A. K. M. Muzahidul Islam, Swakkhar Shatabda*

Department of Computer Science and Engineering, United International University, Bangladesh
Email addresses: {rayhan, salekul, muzahid, swakkhar}@cse.uiu.ac.bd
*Corresponding author: Swakkhar Shatabda (swakkhar@cse.uiu.ac.bd), Department of Computer Science and Engineering, United International University, Plot-2, United City, Madani Avenue, Badda, Dhaka-1212, Bangladesh.

**Abstract**: Precise recognition of emotion from speech signals aids in enhancing human-computer interaction (HCI). The performance of a speech emotion recognition (SER) system depends on the derived features from speech signals. However, selecting the optimal set of feature representations remains the most challenging task in SER because the effectiveness of features varies with emotions. Most studies extract hidden local speech features ignoring the global long-term contextual representations of speech signals. The existing SER system suffers from low recognition performance mainly due to the scarcity of available data and sub-optimal feature representations. Motivated by the efficient feature extraction of convolutional neural network (CNN), long short-term memory (LSTM), and gated recurrent unit (GRU), this article proposes an ensemble utilizing the combined predictive performance of three different architectures. The first architecture uses 1D CNN followed by Fully Connected Networks (FCN). In the other two architectures, LSTM-FCN and GRU-FCN layers follow the CNN layer respectively. All three individual models focus on extracting both local and long-term global contextual representations of speech signals. The ensemble uses a weighted average of the individual models. We evaluated the model's performance on five benchmark datasets: TESS, EMO-DB, RAVDESS, SAVEE, and CREMA-D. We have augmented the data by injecting additive white gaussian noise, pitch shifting, and stretching the signal level to obtain better model generalization. Five categories of features were extracted from the speech samples: mel-frequency cepstral coefficients, log mel-scaled spectrogram, zero-crossing rate, chromagram, and root mean square value from each audio file in those datasets. All four models perform exceptionally well in the SER task; notably, the ensemble model accomplishes the state-of-the-art (SOTA) weighted average accuracy of 99.46% for TESS, 95.42% for EMO-DB, 95.62% for RAVDESS, 93.22% for SAVEE, and 90.47% for CREMA-D datasets and thus significantly outperformed the SOTA models using the same datasets.

## 1. Introduction

The interaction between humans and computers is progressing swiftly. The human-computer interaction (HCI) studies how humans interconnect with computers and to which extent computers are developed to make those human interactions more productive. Interactions between these two entities should be as spontaneous as human-to-human conversations. Therefore, the effective design, proper implementation, and evaluation of interfaces through which the interactions occur are some of the essential focuses of HCI. It aspires to comprehend, assess, and create a range of human experiences, including enjoyment, excitement, concentration, focus, productivity, knowledge, and behavior modification (O'Brien, Roll, Kampen, & Davoudi, 2022). Speech is the principal mode of communication among human beings. Through speech, we humans express one of our most fundamental components, emotions, and the emotion recognition of that speech is one of the active research zones of HCI as well as digital signal processing. The process of distinguishing emotions from speech signals is known as speech emotion recognition (SER). SER is imperative for enhancing the domain of HCI and influential in setting up the direction in which modern-day electronic devices are rapidly moving (J. Chatterjee, Mukesh, Hsu, Vyas, & Liu, 2018). Various significant applications such as intelligent robots, audio surveillance, criminal investigations, automated smart home appliances, movie or music recommendation systems, dialogue systems, etc., which rely on the user's emotional state could do with a system that automatically detects the user's emotion from the speech. Researchers have developed various techniques in the last decade to provide a robust and lightweight SER system (Abbaschian, Sierra-Sosa, & Elmaghraby, 2021; Anvarjon, Mustaqeem, & Kwon, 2020; Khalil et al., 2019). However, due to a lack of technologies and tools, ambiguous nature of emotions, diversity in language and accent across different cultures, frequency and amplitude variation in human utterance regarding gender and age, recognizing the human's emotional states from speech has proven to be complicated and challenging.

A generic high-level overview of the workflow of the proposed SER system is represented in Fig. 1. In the first stage (preprocessing), all the sample audio files are resized to the fixed length, and data augmentation is performed to increase the number of samples and address the data imbalance issues in the datasets by adding AWGN, shifting the pitch, and stretching the time. Next, in the 2$^{nd}$ stage, features from time and frequency domains, as well as commonly used spectral features are extracted from the speech signals. In literature, several studies have derived many features in the area of speech audio processing. Major features types include continuous speech features (e.g., pitch, timing, energy, articulation, and format), voice quality features (e.g., voice level, and phrase), spectral based features such as Mel Frequency Magnitude Coefficient (Ancilin & Milton, 2021), Log-Frequency Power Coefficients (LFPC), Linear-Prediction Coefficients (LPC) (Yusnita, Hafiz, Fadzilah, Zulhanip, & Idris, 2018), and Teager-Energy-Operator (TEO) based features amongst many others (Akçay & Oğuz, 2020). Since these features change over time, speech audio is divided into frames of fitting sizes and low-level descriptor (LLD) features such as MFCC, LMS, ZCR, energy, pitch, Chromagram, tonal centroid features, RMS, RMSE, spectral contrast, centroid, and roll-off are extensively used in the SER task (Anvarjon et al., 2020; Mustaqeem & Kwon, 2020a; Nantasri, Phaisangittisagul, Karnjana, & Boonkla, 2020; Rajamani, Rajamani, Mallol-Ragolta, Liu, & Schuller, 2021). However, the problem of a lower emotion detection accuracy in SER still exists. The features mentioned earlier are presumed to remain constant throughout a frame. As a result, the audio of the speech can be divided into frames, each of which is represented by a feature vector (Ghai, Lal, Duggal, & Manik, 2017). These frames can then be used as a data set for training the proposed models. For this study, ZCR, RMS, Chromagram, LMS, and MFCC features are extracted from the speech audio samples.

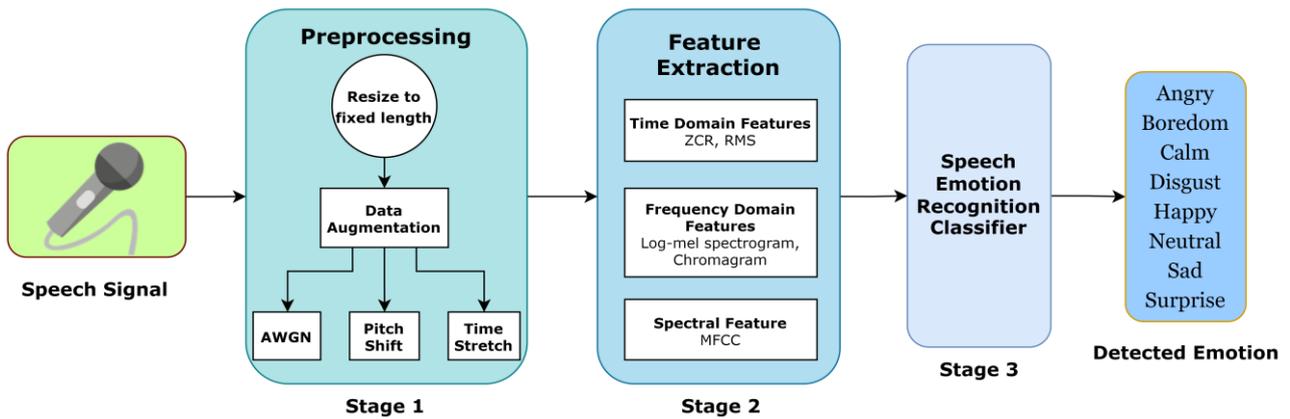

**Fig. 1.** A graphical illustration of the generic workflow of the SER systems.

Previously, the third stage involved classifying features of those speech signals by combining them into feature vectors based on some linear and non-linear classifiers. Frequently used linear classifiers for SER tasks are Support Vector Machine (SVM) (Bhavan, Chauhan, Hitkul, & Shah, 2019), Bayesian Networks (BN) (Ververidis & Kotropoulos, 2008), Linear Discriminant Analysis (LDA) (Z. T. Liu, Xie, et al., 2018), and K-Nearest Neighbors (KNN) (Ezz-Eldin, Khalaf, Hamed, & Hussein, 2021). Due to the non-stationary interpretation of speech signals, it is considered that various non-linear classifiers such as Hidden Markov Model (HMM), and Gaussian Mixture Model (GMM) work resourcefully for the SER task (Venkataramanan & Rajamohan, 2019). However, the recent advancement of deep learning makes it a suitable choice for finding hidden patterns from the extracted feature vectors and performing the classification task.

Deep learning (DL) is a subset of Machine Learning (ML) that has received substantial interest from the research community in recent years. DL has automated the feature extraction process to address the shortcomings of conventional handcrafted feature-based approaches using traditional ML-based methods and enhances the SER task's performance by effectively finding hidden patterns even in the handcrafted extracted features. However, handcrafted features have had much success in SER tasks. Several studies have been carried out in SER utilizing 1D CNN as the classification model (Mustaqeem & Kwon, 2021a, 2021c; Yadav & Vishwakarma, 2020; J. Zhao, Mao, & Chen, 2019). 1D CNN model is effectively used in time-series data and have shown great promise regarding audio classification task. To study the long-term contextual correlations and understand the cues of emotions from speech, researchers have used Recurrent Neural Network (RNN) (D. Li, Liu, Yang, Sun, & Wang, 2021), Long Short Term Memory (LSTM) (J. Zhao et al., 2019), Gated Recurrent Unit (GRU) (Rajamani et al., 2021), and Bidirectional Gated Recurrent Unit (Bi-GRU) (Features, Maji, & Swain, 2022), Bidirectional Long Short Term Memory (Bi-LSTM) (D. Li et al., 2021; Mustaqeem, Sajjad, & Kwon, 2020) along with 1D CNN and Fully Connected Networks (FCNs) (Y. Zhang, Du, Wang, Zhang, & Tu, 2019). Ensemble learning (EL) is a method that refers to the practice of merging several learning models in order to create a more effective and accurate learner. EL has been proven to outperform single estimators. Each estimator in EL is combined in some way, typically by a voting method such as majority voting or weighted voting, to obtain a final result. Numerous studies have proposed using ensemble techniques by combining multiple ML and DL-based models in SER tasks (Chalapathi, Kumar, Sharma, & Shitharth, 2022; Zehra, Javed, Jalil, Khan, & Gadekallu, 2021; Zheng, Wang, & Jia, 2020). However, the existing EL-based studies adopt various traditional

ML-based methods such as SVM (Bhavan et al., 2019), and Random Forest (RF) (Z. Zhang, 2021; Zvarevashe & Olugbara, 2020), or suffers from lower SER rates.

Motivated by the vast success and effectiveness of deep neural networks (DNN) in various classification tasks and higher predictive performance of ensemble learning (EL), in this paper, we propose four DL-based frameworks: first, a baseline dilated 1D CNNs-FCNs based framework; second, a 1D CNNs-LSTM-FCNs based framework; third, 1D CNNs-GRU-FCNs based framework; and fourth an ensemble of those three frameworks through a weighted average mechanism. We have used five publicly available benchmark datasets that are extensively used in the literature: Toronto Emotional Speech Set (TESS) (Pichora-Fuller, Kathleen;, & Dupuis, 2020), Ryerson Audio-Visual Database of Emotional Speech and Song (RAVDESS) (Livingstone & Russo, 2018), Surrey Audio-Visual Expressed Emotion (SAVEE) (Haq & Jackson, 2014), Crowd-Sourced Emotional Multimodal Actors Dataset (CREMA-D) (Cao et al., 2014), and Berlin Database of Emotional Speech (EMO-DB) (Burkhardt, Paeschke, Rolfes, Sendlmeier, & Weiss, 2005). The number of samples in each of these datasets is relatively low for a DL-based model to train properly without any overfitting issues. Besides some of these datasets, have class imbalance issues. To address these challenges, we performed audio data augmentation by injecting additive white gaussian noise (AWGN), changing the pitch of the signal, and stretching the signal level. Additionally, to deal with overfitting we apply kernel and bias regularization (L2=0.01) that reduces the weights squared magnitude, and add dropout layers that arbitrarily eliminate the neurons during the training of the models. The model was trained with augmented data along with the original dataset, yielding a higher accuracy rate with improved generalization ability. Initially, MFCC, LMS, Chromagram, ZCR, and RMS value features are extracted from the audio samples, as previous studies have suggested their efficacy in the SER task (Hajarolasvadi & Demirel, 2019; Lee, Roh, Kim, Kim, & Hong, 2008; Nantasri et al., 2020). The mean value of these features is calculated and is used to train the model to detect human emotions such as "happiness," "sadness," "fearful", "surprise," "anger," "surprise," "boredom," "neutral" etc., from audio signals with improved recognition performance. Combined with data augmentation, each proposed model produces exceptional SOTA results for the SER task. The noteworthy contribution of this work is as follows:

- We propose four DNN-based models built using CNN-based local feature-acquiring blocks (LFABs) and LSTM-GRU-based global feature-acquiring block (GFAB). This work first extracts the LLD features from the speech audio signals.
- The baseline model-A uses seven sequential LFABs to better understand the high-level hidden local features from those extracted LLD features during model training followed by Fully Connected Network (FCN) layers and a softmax layer for classification. The other two models, model-B and model-C, are proposed by adding a GFAB after the final LFAB. Model-B employs LSTM-FCNs, and model-C utilizes GRU-FCNs architecture to acquire long-term global contextual representations from the speech signals. A weighted ensemble framework (model-D) is also proposed which combines the three individual models by adjusting their weights and achieves better performance than the individual models (i.e., model-A, B, and C).
- We have extensively experimented with the models on five widely used publicly available benchmark datasets for SER: TESS, RAVDESS, SAVEE, CREMA-D, and EMO-DB, covering two languages: English and German.
- Data augmentation is performed to increase the training samples, reduce the overfitting problem and make the models more generalized. SER accuracy increased by 3% to 32% from the model trained with the original dataset only.
- The performance of all the proposed models is compared with the previous SOTA models. Amongst all four models, the ensemble model-D achieves the SOTA weighted average accuracy of 99.46% for TESS, 95.42% for EMO-DB, 95.62% for RAVDESS, 93.22% for SAVEE, and 90.47% for CREMA-D datasets. These are significantly improved results compared to the single models and the previous SOTA methods on each dataset.

The remainder of this paper is assembled as follows. Section 2 presents the existing literature review in the SER task to grasp the current trend, intuition getting, and find the scope for improving the task. Section 3 provides an overview of the architecture of the proposed models. Section 4 is covered by an in-depth discussion about utilized datasets, data augmentation techniques, the feature extraction process, and model training. We comprehensively analyze and compare the experimental results of the proposed individual model-A, B, C, and weighted ensemble model-D with SOTA SER benchmarks in section 5. Finally, in section 6, we conclude with a discussion about the existing challenges and possible future research directions in SER.

**2. Related Works**

Digital signal processing (DSP) is a matter of great interest among the research community, and researchers have come up with several approaches for a robust improvement by eliminating the existing issues in the SER task. For any SER task, feature selection is the most critical part because irrelevant features directly affect the next part, which is speech emotion classification. Currently, researchers worldwide are utilizing DL for SER-related tasks due to their vast triumphs in representing features and the ability to find hidden patterns from the extracted speech-based feature set. In time-domain representation, some of the commonly utilized

SER features are amplitude envelope, ZCR, and RMS (Das et al., 2022). Notable frequency domain-represented speech features include band energy ratio, Mel-scaled spectrogram, Chromagram, and various spectral features such as centroid, flux, contrast, and roll-off (Alnuaim, Zakariah, & Alhadlaq, 2022). The most significant and widely used cepstral-based feature for the SER task is MFCC. Statistical features include entropy, skewness, kurtosis, etc. Recently gammatone cepstral coefficients (GTCC) are being heavily explored in the field of SER (Bandela & Kumar, 2021; S. Zhao, Yang, Cohen, & Zhang, 2021). Speech spectrogram is one of the significant features utilized by most researchers nowadays regarding SER tasks (Alnuaim et al., 2022; Mustaqeem & Kwon, 2021b; Sultana et al., 2022). It is a two-dimensional (2D) depiction of speech signals. It visually represents a signal's power, where different frequencies over time are shown in the waveform. MFCC is the most widely used feature in terms of SER tasks. By converting the traditional frequency to mel-scale, MFCC accounts for human insight for sensitivity at acceptable frequencies, making it appropriate for SER tasks. When training models, 12 to 20 MFCCs are usually taken into account, containing information about the changes in rate in the different spectrum bands (Abdel-Hamid, 2020; Christy, Vaithyasubramanian, Jesudoss, & Praveena, 2020; Issa, Fatih Demirci, & Yazici, 2020; Nantasri et al., 2020). The ZCR feature represents the positive and negative sign changes in a signal (J. Chatterjee et al., 2018). Chroma-based audio features have proven effective for investigating, evaluating, and extracting information from music audio and SER-related tasks (Issa et al., 2020). It is observed that methods such as traditional ML, DL, and their fusion are extensively used in recent literature for the SER task. The researchers are also exploring several attention mechanisms along with DL methods to focus more on the region of interest in the speech signal. Aside from these mentioned methods, ensemble learning and transfer learning-based methods are gaining momentum due to their increased performances in various classification tasks.

*2.1 Traditional machine learning-based models*

Numerous methods based on SVM, Multi-Layer perceptron (MLP), Gaussian Mixture Modelling (GMM), Naïve Bayes (NB), and Hidden Markov models (HMM) were utilized in earlier efforts for SER from the speech signal. (Ancilin & Milton, 2021) use the SVM classifier for the SER task on SAVEE, EMO-DB, RAVDESS, eNTERFACE, EMO-VO, and Urdu datasets. The magnitude spectrum was employed instead of energy spectrum to extract the Mel frequency magnitude coefficient features from the speech signals. In addition, traditional MFCC, log frequency power coefficient, and linear prediction cepstral coefficient were also extracted and utilized for model training. SER performance is enhanced by using the magnitude spectrum instead of the power spectrum and using the log magnitude coefficients directly rather than the cosine transformed coefficients. (Z. T. Liu, Wu, et al., 2018) propose an SER framework that selects features based on correlation analysis and Fisher criterion. This process reduces the number of irrelevant and redundant features. Then they used the extreme- ML decision tree scheme to classify the emotions into different categories by utilizing the Chinese emotion corpus, Chinese Academy of Science Institute of Automation (CASIA). (Wang, An, Li, Zhang, & Li, 2015) suggest a new kind of speech feature, Fourier parameter functions along with the first and second-order differences of harmony-based features, estimated by Fourier analysis. They have extracted and combined two types of features: Fourier parameter and MFCC from the utilized EMO-DB, CASIA, and Chinese elderly emotion database (EESDB) datasets and employed the combined features as input to the SVM and a Bayesian classifier for the SER task. (Palo, Chandra, & Mohanty, 2017) recommend an SER technique using MLP and GMM for the Oriya Language. They used various feature extraction techniques, including MFCC, Perceptual linear prediction, and linear predictive coding. MLP achieves the highest SER accuracy of 87%. (Demircan & Kahramanli, 2018) utilize type-1 fuzzy C-means method to the extracted MFCC and linear prediction coefficient features and after that identified the cluster centers and fed them to different classifiers such as SVM, KNN, and NB for the classification task. The SVM classifier achieved the highest classification rate of 92.86%. However, there exist some issues with methods like HMM and GMM, for example, finding the most likely sequence of hidden states, given a sequence of observations. Another significant disadvantage of these methods is that they underperform while modeling nonlinear data (Venkataramanan & Rajamohan, 2019).

*2.2 Deep learning-based models*

Deep learning (DL) methods such as ANN, DNN, CNN, RNN, GRU, LSTM, Bi-LSTM, and Bi-GRU have been leveraged as feature extractors to facilitate the learning of discriminative representations, with varying degrees of success (Anvarjon et al., 2020; Mustaqeem et al., 2020; Yadav & Vishwakarma, 2020; S. Zhang et al., 2020). (Nantasri et al., 2020) propose an SER model by collecting 20-MFCC, 20-delta, and 20-delta-delta features and computing their mean values. These mean values are used as the input for the artificial neural network (ANN) classifier. They have evaluated their model with RAVDESS and EMO-DB datasets and achieved 82.3% and 87.8% accuracy, respectively. To reduce the error rate of ANN and proper selection of optimal weights and biases for the model to train, (Moghanian, Saravi, Javidi, & Sheybani, 2020) propose a new technique named GOAMLP. (Lalitha, Tripathi, & Gupta, 2019) propose a deep neural network (DNN) model for the SER task to investigate the effective predictive performance of perceptual-based speech features. (Anvarjon et al., 2020) propose an SER model based on extracting high-level features from the spectrograms of speech utterances. They have used plain rectangular kernels with a revised pooling strategy. The model's performance was evaluated with two datasets. It achieved 77.01% and 92.02% accuracy for the Interactive Emotional Dyadic Motion Capture (IEMOCAP) and EMO-DB datasets, respectively. (Yoon, Byun, & Jung, 2018) implement a dual recurrent encoder model approach for SER tasks by utilizing text and audio data from the IEMOCAP dataset. The authors employ MFCC derivatives and prosodic features along with text tokens as the input features of the proposed framework. Their multimodal method led to an accuracy of 71.8% on the IEMOCAP dataset. (Tiwari, Soni, Chakraborty, Panda, & Kopparapu, 2020) propose an utterance-level parametric generative noise model to test the robustness of the SER model when exposed to the presence of additive noise. Their proposed architecture is advantageous for suppressing unseen noise because the manufactured noise can encompass the total noise space in the energy domain of the Mel-filter bank. However, even with the performed data augmentation (DA), the achieved DNN-based SER model's performance is not very significant, with an accuracy of 76.77% on the EMO-DB dataset and 53.35% on the IEMOCAP dataset. (Neumann & Vu, 2019) integrates unsupervised auto-encoder strategy along with the CNN method to classy emotions from speech, however, this unsupervised approach achieves less satisfactory performances in terms of SER. Recently unsupervised DL-based algorithms are being explored for the SER task due to the shortage of sufficient data samples in each of the publicly available datasets. DA through unsupervised DL-based algorithms such as generative adversarial networks (GAN) (Chatziagapi et al., 2019), conditional GAN (Ma, Li, Ni, Huang, & Zhang, 2022), cycle consistent GAN (Bao, Neumann, & Vu, 2019) is performed in many studies, however, one noticeable thing is, the use of completely synthetic data in those studies achieve unsatisfactory performance regarding SER task. (Praseetha & Joby, 2021) employed a GRU-based DL model for the SER task and extracts the filter-bank energies of the speech signals to train the model. The model achieves an accuracy of 93% in the augmented TESS dataset. In another work of SER, (Jothimani, S and Premalatha, 2022) utilize the CNN and LSTM-based models where the experimental analysis was carried out using the SAVEE, CREMA, RAVDESS, and TESS datasets. The authors used the MFCC, ZCR, and RMS value as the features for model training.

*2.3 Hybrid models*

Due to the success of these DL-based architectures, interest in fusing these network types into a single architecture to capture both local and long-term contextual dependencies of data has increased recently (S. Li et al., 2021; U. Kumaran, Rammohan, Nagarajan, & Prathik, 2021; Xu, Zhang, & Zhang, 2021), ensemble learning (Chalapathi et al., 2022; Zheng et al., 2020) make up most SER architectures that use neural networks. (Sultana et al., 2022) performs a cross-lingual SER study by utilizing CNN and the Bi-LSTM network that tries to capture both temporal and sequential representations of emotions. (Mustaqeem & Kwon, 2020b) extract spatiotemporal features for the SER task using a ConvLSTM model. Using four blocks of 1D CNN and LSTM, the authors have gathered the most significant distinctive emotional features. The extracted features are then fed into the GRU-based network, which is used to re-adjust the global weights. By utilizing the 1D CNN with LSTM network in one model and 2D CNN with LSTM network in another model, (J. Zhao et al., 2019) propose two SER models. The 2D CNN LSTM model achieves better emotion recognition results by focusing on capturing local correlations as well as global contextual information from LMS features. Researchers have experimented with traditional ML-based methods and DL-based methods in the same work and did a comparative analysis of these models' SER performance. (Singh, Puri, Aggarwal, & Gupta, 2020) leverages the CREMA-D dataset to train two classifiers (SVM and RNN) with prosodic and spectral features, that account for variance in speech intensity. The classifiers were trained at three stages of intensity: low, medium, and high. The "Happy" and "Neutral" labeled emotions have the highest classification accuracy, while the "Disgust" labeled emotion has the lowest. (Kerkeni et al., 2019) suggest an automatic SER system based on machine learning methods. The authors extract modulation spectral and MFCC features from speech signals in two corpora of EMO-DB and Spanish speech utterances and classifies them using SVM, Multivariate linear regression (MLR), and RNN classifiers. Feature selection was used to identify the most relevant feature subset. SER reported the highest recognition rate of 94% using the RNN classifier without speaker normalization and feature selection on the Spanish dataset.

*2.4 Attention-based models*

Recently different attention mechanisms are being extensively integrated into the SER domain due to their ability to explore distinctive regions of data. (Guo et al., 2022) integrate phase information of a signal with the magnitude information for the SER task. To capitalize on the complementary nature of magnitude and phase information, this study employs a single-channel model along with a multi-channel model with attention based on magnitude spectrograms, modified group delay cepstral coefficients, and dynamic relative phase. Incorporating phase information makes it possible to capture more comprehensive acoustic features. (Z. Zhao et al., 2021) propose a hybrid deep CNN architecture that leverages parallel convolutional layers combined with a squeeze-and-excitation network incorporated with a self-attention-based dilated residual network. The architecture is trained with connectionist temporal classification loss for discrete SER tasks and effectively captures long-term contextual dependencies. To investigate the autocorrelation of phonemes in speech, (D. Li et al., 2021) combine the self-attention mechanism with the Bi-LSTM network. The self-attention mechanism can provide different weights to frames of varying emotional intensity, but it can also determine the autocorrelation between frames. (S. Li et al., 2021) propose a composite model that combines a spatiotemporal attention network, with a frequency-based attention network. The proposed network narrows down the emotional frequency regions from a spectrogram image to focus on the desired emotional regions. They have also developed a large-margin learning technique to deal with the problem of feature aliasing. It improves intra-class compactness while increasing inter-class distances among features. (G. K. Liu, 2018) demonstrates that a feature set consisting of gammatone frequency cepstral coefficients improves the SER accuracy by 3.6% over MFCCs by investigating three frameworks: Fully Connected Networks (FCN), LSTM, and Attention-LSTM networks. (Yoon et al., 2019) present a multi-hop attention framework for the SER task by extracting hidden contextual information from speech data using two streams Bi-LSTMs and then applying the multi-hop attention strategy to generate the final weights for emotion recognition. (Meng, Yan, Yuan, & Wei, 2019) propose a 3D LMS-based residual dilated CNN and memory attention mechanisms. They utilize a composite of static LMS feature, delta, and deltas-deltas feature to build the feature vector from the raw speech signal as input for the model. The dilated CNN assists the model to obtain more receptive fields than using the conventional pooling layer. In the IEMOCAP (speaker-dependent) dataset the model achieved 74.96% accuracy, and in the IEMOCAP (speaker-independent) dataset it achieved 69.32% accuracy. The model achieved the best accuracy of 90.37% on the EMO-DB (speaker-dependent) dataset. (Mustaqeem & Kwon, 2021b) designed a self-attention module based on DL for the SER system. It receives the transitional feature maps and uses it to build the channel and spatial attention map with minimal overhead. The authors employ a dilated CNN architecture in spatial attention to extract spatial information and a multi-layer perceptron in channel attention to extract global cues from the input tensor. The proposed model archives 78.01%, 80.00%, and 93.00% accuracy on IEMOCAP, RAVDESS, and EMO-DB datasets, respectively. (Xie et al., 2019) propose an SER system based on modified attention-LSTM architecture. They have extracted frame-level speech features from the waveform to replace traditional statistical features, preserving the timing relations in the original speech through the sequence of frames. The forget gate of the LSTM was replaced with an attention gate in order to reduce complexity. Additionally, they increased the system's efficiency by applying the attention mechanism on both time and feature dimensions rather than simply forwarding the previous iteration's output in LSTM. Although attention modules have become an integral component of modern SER systems, they are not indispensable for achieving high SER performances or even SOTA results.

*2.5 Transfer learning-based models*

Methods based on the use of pre-trained neural networks frequently produce superior performances to more traditional procedures. Transfer learning (TL) has the potential to overcome SER's cross-domain barrier. (S. Zhang, Zhang, Huang, & Gao, 2018) employed pre-trained AlexNet architecture (Krizhevsky, Sutskever, & Hinton, 2012) for learning high-level feature representations from the extracted three channels of the LMS feature. Additionally, the authors suggest an approach for pooling named discriminant temporal pyramid matching (DTPM) features to discriminative utterance-level representations. AlexNet fine-tuned for emotional speech outperformed the simpler Depp CNN model in four distinct datasets, while DTPM-based pooling outperformed the traditional average pooling method. A 2D CNN-based model that uses spectrograms generated from the EMO-DB dataset, (Badshah, Ahmad, Rahim, & Baik, 2017) propose an SER architecture. They have also explored the field of transfer learning and utilized pre-trained AlexNet architecture but got unsatisfactory results. The initial proposed model achieved 84.3% accuracy on the test set. (Xi, Li, Song, Jiang, & Dai, 2019) used a residual adapter to minimize domain-specific parameters while increasing domain-agnostic parameters sharing. (Aggarwal, Srivastava, Agarwal, & Chahal, 2022) propose a two-way feature extraction method for the SER task. In the first approach, they extract the MFCC, spectrogram, spectral centroid, and roll-off features. Then in the second approach, they extract 2-dimensional LMS images from the speech signals. They utilize the pre-trained VGG-16 network for the SER task.

*2.6 Ensemble learning-based models*

Ensemble Learning (EL)-based methods have higher predictive accuracy compared to individual estimators. It combines the predictions from two or more ML or DL models to produce a more stable, accurate, and robust prediction. (Chalapathi et al., 2022) utilize the adaptive boosting ensemble method along with the fuzzy c-means approach to deal with the high-dimensional acoustic features. (Zvarevashe & Olugbara, 2020) employ bagging classifiers such as random decision forest, bagging with SVM, MLP, and boosting classifiers such as gradient boosting machine, and AdaBoost with CART for the SER task. In another study, (Z. Zhang, 2021) used the RF classifier along with the weighted binary cuckoo search method to select the optimal feature subset. Though time required to train multiple ML and DL-based architectures to perform the ensemble mechanism for the classification task is still a matter of concern. However, emotion is a sensitive topic, and recognition of emotion from the speech is a challenging task. A combination of multiple individual estimators and utilizing each of their feature learning strengths in the SER domain using an ensemble mechanism even at the cost of larger training time should be considered because of the need for more accurate and stable recognition performance of speech emotions. Inspired by the efficient and stable predictive performance of EL-based architecture, we adopt the EL mechanism for the final model-D, which combines the predictive results of three proposed individual models–A, B, and C in a weighted average method.

In Tables 4 to 9, we provide an extensive comparative evaluation between our proposed work and the notable works discussed above in the literature review section. In the comparison, we highlight the methodology, extracted features, utilized datasets, feature dimension, data augmentation methods, achieved results, and year of publications.

## 3. Proposed Methods

The main objective of this work is to investigate the efficiency of an ensemble architecture combining multiple novel DL-based models for a multi-lingual SER system. Factors such as utilized datasets, number of samples in those datasets, class imbalance, data augmentation (DA), feature extraction from speech signals, and selection of proper ML or DL-based classifiers play a significant role in the SER performance of an ensemble architecture. Fig. 2. summarizes our approach to an ensemble architecture for the SER task. This study utilizes five benchmark SER datasets (TESS, EMO-DB, RAVDESS, SAVEE, and CREMA-D) covering English and German languages. Since there is a shortage of sufficient sample audio files in those datasets, and to deal with the adverse impact of this data shortage issue on the performance of DL-based architecture, we performed three types of DA techniques (AWGN addition, pitch shifting, and time stretching) to increase the samples of those datasets to obtain proper convergence and generalizability of the proposed DL-based models. The proposed approach involves extracting a combination of time-domain, frequency-domain, and cepstral-based features from raw audio recordings to provide the DL-based models as input. The predictions from the individual DL-based models are then weighted, and a weighted average ensemble prediction is performed with SOTA SER performance. We present further details about the utilized datasets, adopted DA techniques, and extracted speech features in section 4. The details of the proposed individual DL model-A, B, C, and ensemble model-D are presented below.

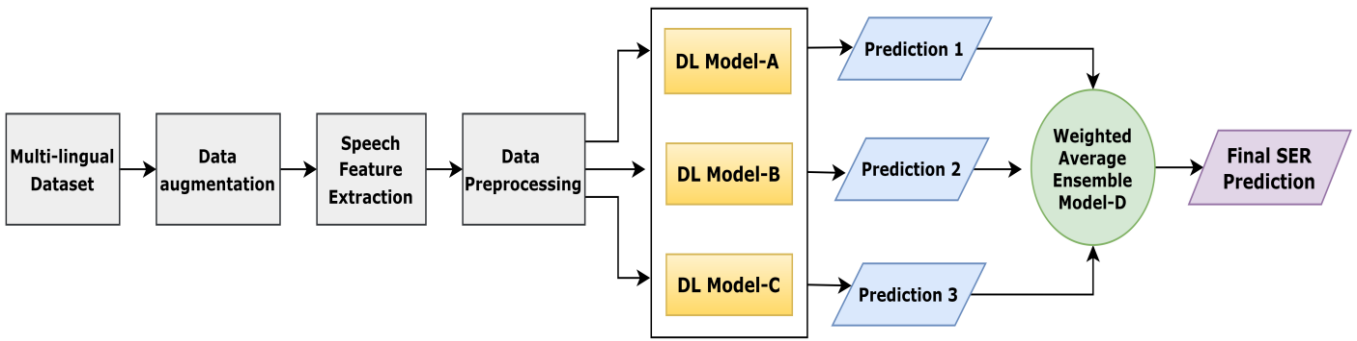

**Fig. 2.** A graphical representation of the framework of the proposed ensemble architecture.

### 3.1. Proposed baseline model-A (1D CNNs-FCNs)

This study uses 1D CNN followed by FCNs to build the first baseline model for SER. 1D CNN performs well with structured data. In terms of audio data, 1D CNN extracts the temporal information within the speech signal. The extracted ZCR, Chromagram, MFCC, RMS, and LMS features from the speech signals are stored in an array creating a vector of features. This vector of features is fed to the proposed baseline model as input. Using seven sequential LFABs containing convolutional, max-pooling, batch normalization (BN), and dropout layers, the model extracts hidden local patterns from the speech audio signals, as shown in Fig. 3. The 1D-convolution layer, max-pooling layer, and BN layers are the essential layers of the LFABs. Two FCNs collect the ultimate global features from the speech signals.

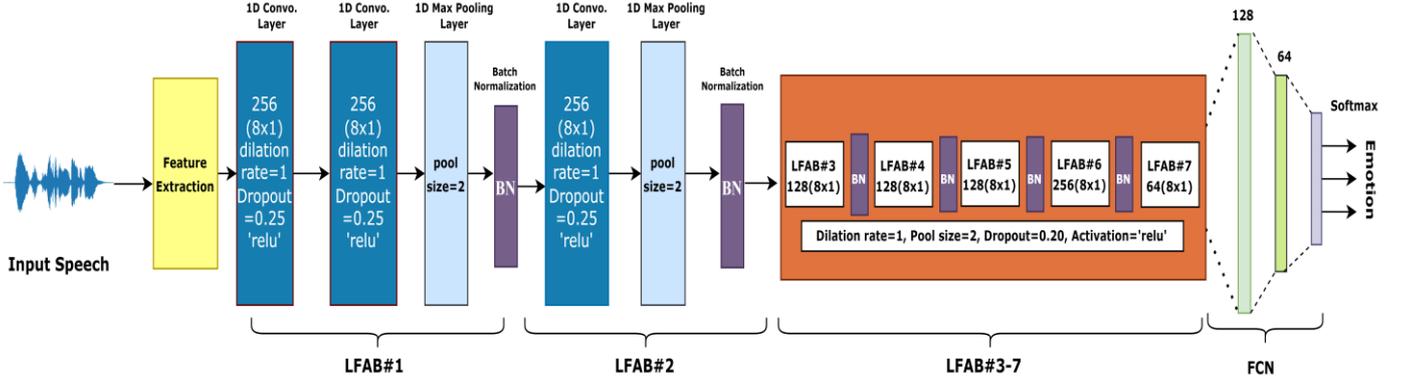

**Fig. 3.** The architecture of the proposed baseline model-A (1D CNNs-FCNs).

The proposed 1D CNNs-FCNs Model-A takes 155x1 feature vector arrays as the input. The first LFAB block has 256 filters, with a kernel size of eight, padding = 'same', dilation rate = (1x1), and a stride of one. The dilation rate reduces the input vector feature map. The Rectified Linear Unit (ReLU) then triggers its output after BN is added. By solving the vanishing gradient problem, the BN layer aids all layers of the neural network in learning at a normalized rate. It speeds up the training process by normalizing the hidden layer activation. In addition, to cope with the model overfitting issue, we used the dropout layer and kernel regularization (L1 and L2) methods with a rate of 0.01. The output of a preceding input layer is received by the second layer in this stack, which consists of identical 256 filters with the corresponding kernel size, dilation rate, and stride. ReLU also enables the output of this layer, and then dropout at a rate of 0.25 is added. Following that, BN is performed, with the output being fed to a 1D max-pooling layer with a window size of two. The following six LFABs with filters of 256, 128, 128, 128, 256, and 64 filters use the kernel size, dilation rate, and stride configuration as previous blocks. The flattening layer and 50% dropout follow the ultimate LFAB. This flattening layer output is received by two FCNs of 128, and 64 units with a dropout of 50%, and finally, the output layer with softmax activation function which is utilized to distinguish the emotion according to the hidden features learned through LFABs. Depending on the task, the LFABs can be customized differently. The changes in LFAB configuration are primarily reflected in the convolution, dilation, pooling, and batch normalization settings.

*3.2 Proposed model-B (1D CNNs-LSTM-FCNs) and model-C (1D CNNs-GRU-FCNs)*

Proposed model-B and model-C, as shown in Fig. 4 and 5 respectively, are built on top of model-A. Here, we see that after the final LFAB in baseline model-A, one additional global feature acquiring block (GFAB) comprising of LSTM layer (model-B), and GRU layer (model-C) of 512 units is added to learn the global contextual correlations from the features engineered through the LFABs, as well as adjusting the global weights. GFAB is followed by a dropout layer of 50%. The FCNs configuration remains the same as model-A. We adopt GRU and LSTM architecture to obtain global long-term contextual representations in speech utterances. In an LSTM cell as shown in Fig. 4, there are three gates: forget, input, and output gate. The three gates control the transfer of information into and out of the cell, and the cell retains values over different periods. Gates are a way to allow information to pass through selectively.

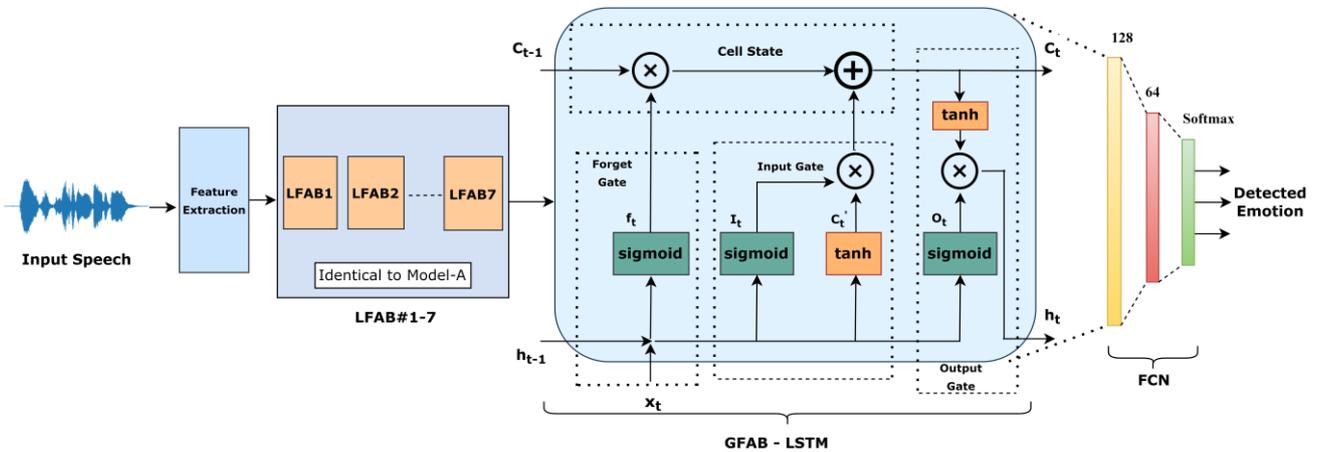

**Fig. 4.** The architecture of the proposed model-B (1D CNNs-LSTM-FCNs).

The forget gate ($f_t$) determines what essential information is to retain from the preceding cell state ($C_t$) and can be calculated using Eq. (1). The input gate ($I_t$) defines what pertinent information can be incorporated from the current time step, while the output gate ($O_t$) defines the current hidden state that will be sent to the subsequent LSTM unit. The $I_t$ and $O_t$ calculation formula is shown in Eq. (3) and (4) respectively. First, a sigmoid layer determines which value to update. After that, to regulate the network, a hyper tangent (*tanh*) layer generates a feature vector $C'_t$, with possible values between -1 and 1. The following step updates the information from the previous cell state to the new cell state through Eq. (5). Usually, the length of the feature of frame-level speech changes with the number of speech frames. The LSTM learns deep global contextual features with fixed length by choosing the output of the last timestep from the variable-length frame-level speech features (Xie et al., 2019). Finally, the output is calculated using Eq. (6) and (7) (J. Zhao et al., 2019).

$$f_t = \sigma(W_f * [h_{t-1}, x_t] + b_f) \tag{1}$$

Where $h_{t-1}$ is the hidden later output at the previous timestep, $x_t$ is the current timestep input, $W_f$ is the weight matrix between $I_t$ and $O_t$, and $b_f$ represents the connection bias at timestep $t$. $\sigma$ is the logistic sigmoid function which is calculated using Eq. (2).

$$\sigma(x) = \frac{1}{1+e^x} \tag{2}$$

$$I_t = \sigma(W_i * [h_{t-1}, x_t] + b_i) \tag{3}$$

$$C'_t = \tanh(W_C * [h_{t-1}, x_t] + b_C) \tag{4}$$

$$C_t = f_t * C_{t-1} + I_t * C'_t \tag{5}$$

$$O_t = \sigma(W_o * [h_{t-1}, x_t] + b_o) \tag{6}$$

$$h_t = O_t * \tanh(C_t) \tag{7}$$

Where, $W_i$, $W_C$ are the weight matrices between $O_t$, $I_t$, and $\sigma$ respectively. $b_C$ represents the bias vector for $W_C$, $C_t$, $C_{t-1}$, $f_t$, $I_t$ represents the information of cell state, previous timestep, forget gate, and input gate at timestep $t$ respectively. The value generated by *tanh* is $C'_t$, whereas, $W_o$ and $b_o$ represents the weights and bias of the $O_t$ at timestep $t$.

The GRU is similar to the LSTM. It only has one hidden state compared to LSTM's two states: cell and hidden (Chung, Gulcehre, Cho, & Bengio, 2014). Due to the gating mechanisms, this hidden state can hold both long-term and short-term dependencies simultaneously. As shown in Fig. 5, the GRU cell is a combination of two gates: update, and reset gates, but the internal structure is different from LSTM. While training, the gates learn what information is essential to retain or overlook. The update gate in the GRU replaces the forget and input gates of the LSTM. Reset gate aid in the capture of the sequence's momentary representations. The reset gate, update gate, candidate hidden state, and the final hidden state of the GRU can be calculated through the following equations (8) to (11) (Ravanelli, Brakel, Omologo, & Bengio, 2018).

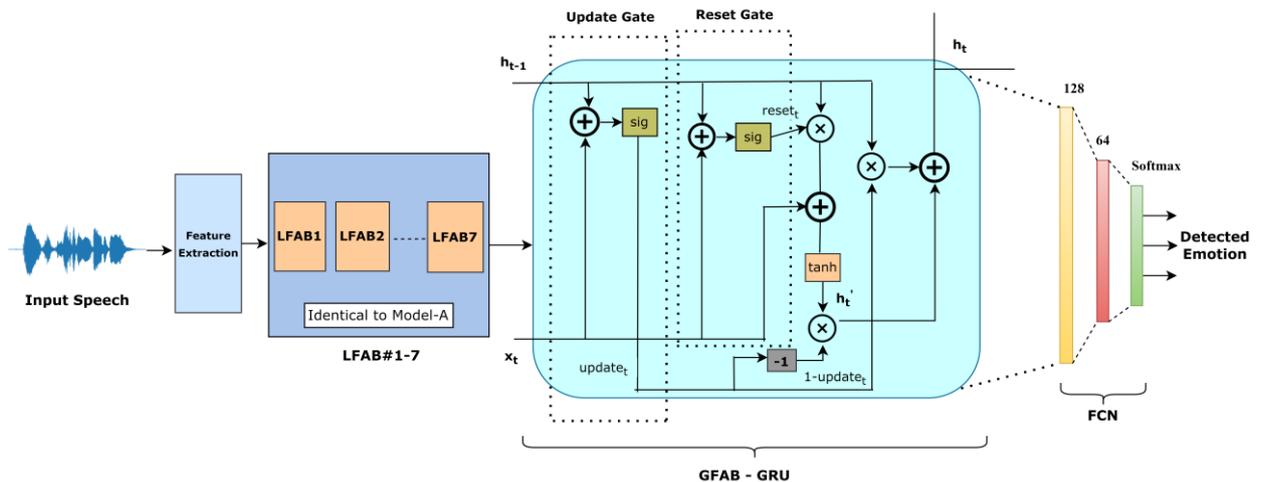

**Fig. 5.** The architecture of the proposed model-C (1D CNNs-GRU-FCNs).

The output of the reset gate is obtained by multiplying the preceding hidden state $h_{t-1}$ and current timestep input $x_t$ by their respective weights, adding them, and then applying a sigmoid function to the sum. The update gate $update_t$ differs from the reset gate only in terms of the weight metrics. The candidate hidden state or current memory function $h'_t$ is then calculated by multiplying the input vector $x_t$ with $h_{t-1}$ and then performing an element-wise multiplication with the reset gate, $reset_t$. This $h'_t$ is then used to calculate the final hidden state $h_t$.

$$reset_t = \sigma(W_{reset} * [h_{t-1}, x_t] + b_{reset}) \quad (8)$$

$$update_t = \sigma(W_{update} * [h_{t-1}, x_t] + b_{update}) \quad (9)$$

$$h'_t = \tanh(W_h * [reset_t \odot h_{t-1}, x_t] + b_h) \quad (10)$$

$$h_t = (1 - update_t) \odot h_{t-1} + update_t \odot h'_t \quad (11)$$

Since GRU has fewer gates than LSTM, it is less complicated and faster to train. GRU should be used if the dataset is relatively small; otherwise, LSTM should be applied for large-volume datasets.

*3.3 Proposed weighted ensemble model-D*

Ensemble learning (EL) combines the learning procedures of several models to achieve a more stable and comprehensive prediction with a maximum accuracy that is superior to the individual DL models' accuracy. Specific models are good at modeling one part of the data, and others are good at modeling another. EL succeeds because several models will not make identical errors in the same test dataset. It assures that the most accurate and reliable prediction is generated. Many features in the field of SER can reflect the emotion of speech. When the distinct advantages and accuracy of various SER-related models are merged and the features are combined, the recognition efficiency can be significantly enhanced. A weighted average ensemble was performed in this study by combining model-A, B, and C (see Fig. 6). At first, we select the optimal weights for each of the individual models through the Grid-Search technique. Then using the tensordot function of NumPy, we multiply the selected optimal weights with the prediction results of each model, calculate the sum of this product of elements over the specified axis to calculate the weighted prediction result, and then find the class with the largest predicted probability. Then from this weighted prediction result, the class with the largest predicted probability is chosen for the final prediction. The weighted average ensemble model-D, tested with the original dataset and augmented data, achieved higher weighted average accuracy (WAA) than the individual models- A, B, and C.

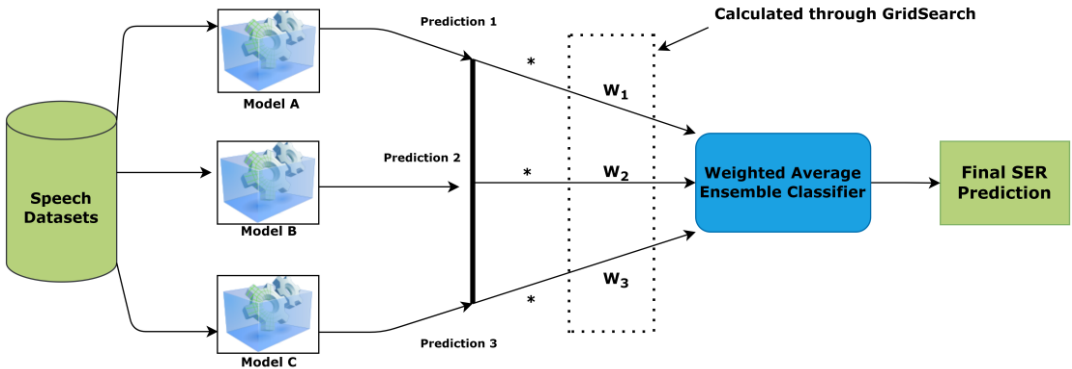

**Fig. 6.** Visual representation of the proposed weighted ensemble model-D.

**4. Experimental Analysis**

Primarily, an SER framework is comprised of two components. The first one is the preprocessing component that obtains appropriate features from the speech utterances of the utilized datasets, and the second one is a classifier that uses those obtained features to execute the SER task. This section provides details about the datasets utilized in this study, data augmentation techniques, extracted features, and model training.

*4.1. Datasets*

To perform a meticulous evaluation of the proposed models, five datasets were explored covering two languages: English and German. We performed data augmentation in all the datasets since the number of samples in each of them is not significant for a DL-based model to train appropriately. A summary of each of them is provided below.

**Toronto Emotional Speech Set (TESS):** TESS (Pichora-Fuller et al., 2020) is the first dataset that this study explored. It contains 200 target words. Those words were spoken by two English actresses, ages 26 & 64 years, respectively. The dataset is well balanced and contains 2800 audio files and depicts seven emotions: "angry," "neutral," "happy," "disgust," "surprise," "fear," and "sad". Note that this dataset has not been extensively used in SER studies previously. After performing data augmentation, the samples increased to 8400 samples. The average sample duration for all datasets is 2.8 seconds, with TESS being the outcast with an avg. period of 2.1 seconds.

**Ryerson Audio-Visual Database of Emotional Speech and Song (RAVDESS):** RAVDESS (Livingstone & Russo, 2018) is one of the most explored datasets in the SER tasks. It includes both audio and video recordings of twelve male and twelve female actors reciting English sentences while exhibiting eight distinct emotional expressions. For this study, only the speech audio samples were utilized. The total number of audio files is 1440 with a sampling rate of 48 kHz, with 60 trials per actor. Only the speech audio sample from the dataset of the following eight categories are covered in this study: "sad," "happy," "angry," "calm," "fearful," "surprised," "disgust," and "neutral". It is a balanced dataset though the "neutral" class has a smaller number of records compared to other classes. After performing data augmentation, the samples increased to 7200.

**Surrey Audio-Visual Expressed Emotion (SAVEE):** SAVEE (Haq & Jackson, 2014) consists of 480 speech utterances spoken by four English actors aged 27 to 31 years in seven diverse emotions: "angry," "happy," "neutral," "disgust," "sad," "fear," and "surprise" in a phonetically stable manner. The utterances are sampled at a rate of 44.1 kHz with a resolution of 16 bits However, this dataset has a class imbalance issue, with the "neutral" class being almost double compared to all the other classes. For this study, only the speech audio samples were utilized. Data augmentation increased the samples to 1920.

**Berlin Database of Emotional Speech (EMO-DB):** EMO-DB (Burkhardt et al., 2005) is the most well-known and extensively used dataset in the SER research field. The utterances are sampled at a rate of 16 kHz with a resolution of 16 bits. It comprises 535 audio recordings in the German language categorized into seven emotional kinds: "anger," "fear," "sadness," "happiness," "disgust," "boredom," and "neutral". However, this dataset has a class imbalance issue, with the "anger," class utterance number being large compared to other classes. With data augmentation, the samples increased to 2140.

**Crowd-Sourced Emotional Multimodal Actors Dataset (CREMA-D):** CREMA-D (Cao et al., 2014) is the least explored dataset in the SER research field. It uses 7442 recordings from ninety-one actors/actresses (48 male actors and 43 female actresses) from diverse races and customs, making it the most complicated to use. Actors spoke from a group of twelve sentences of six different emotional categories: "angry," "happy," "neutral," "disgust," "fear," and "sad". Though the original number of samples is quite large compared to the other four datasets, it is still considered one of the most challenging datasets to work with because of its diverse number of male and female speakers. Data augmentation was also performed in this dataset with increased samples of 44652. Fig. 7 shows the number of the class-wise utterance of each of the datasets.

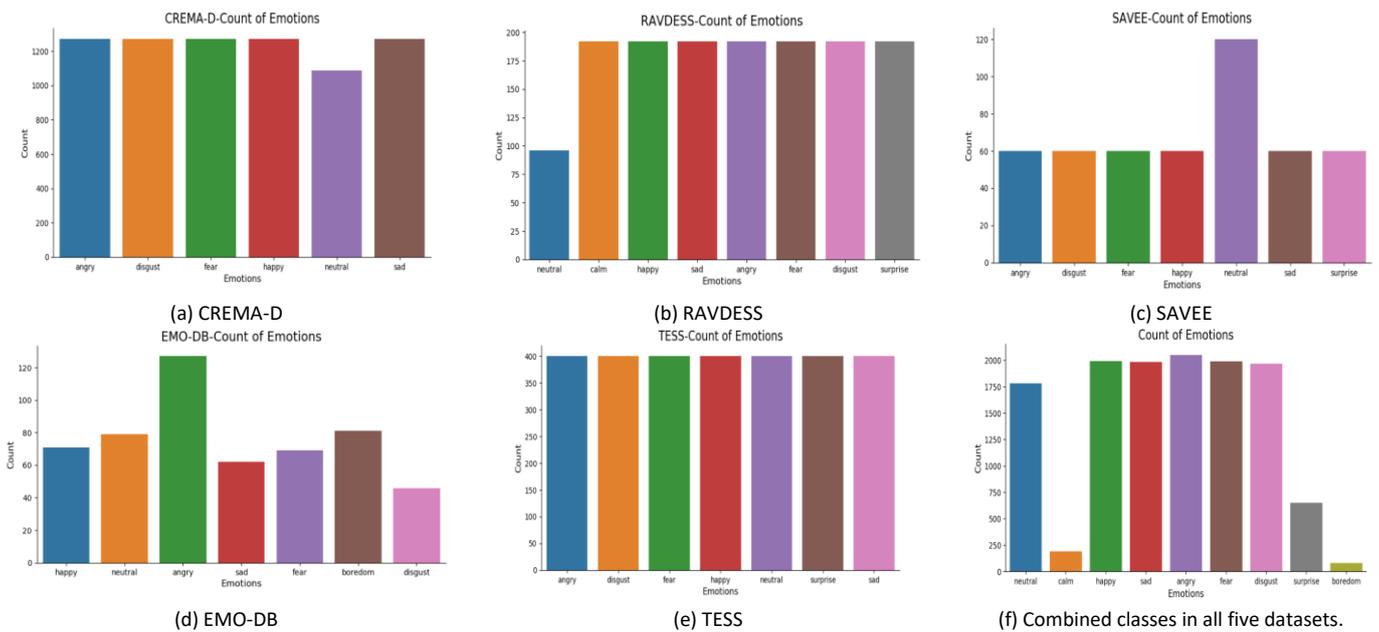

**Fig. 7.** Class-wise utterance distribution in all five datasets, (a) CREMA-D, (b) RAVDESS, (c) SAVEE, (d) EMO-DB, (e) TESS, and (f) Combined.

*4.2. Data augmentation*

Frequently observed issues in the SER task include the insufficient size and class imbalance of datasets. As the complexity and scale of DNNs expand, a substantial dataset is required for their optimal performance. One solution is to increase the dataset using diverse data augmentation (DA) techniques. DA is the method of applying minor modifications to our original training dataset to produce new artificial training samples. Since the number of speech utterance records in each class is relatively low, this study performs three types of audio DA, additive white gaussian noise (AWGN) injection, time-stretching, and pitch shifting in the audio files. The impact is more data for proper training of the models. The impact of these techniques is visually presented in Fig. 8. The obtained signal with AWGN is equal to the transmitted signal with some added noise, which is statistically independent of the signal. AWGNs are random samples dispersed at consistent intervals with a mean value of zero and a standard deviation of one. We added AWGN to the samples by using NumPy's normal and uniform method with a rate of 0.020, and 0.025. We can adjust the speed or duration of a sound sample without changing the pitch by stretching time. We performed this task by using the *time_stretch* method of python's librosa library, with a factor of 0.7 and 0.8. We also changed the sound's pitch without affecting the speed. Pitch shifting was done by using the *pitch_shift* method of librosa, with a factor of 0.6 and 0.7. Several other studies have performed DA for the SER task using GAN-based methods (Bao et al., 2019; Shilandari, Marvi, Khosravi, & Wang, 2022; Tiwari et al., 2020). However, those augmentation methods did not yield higher SER performances. We present a comparative analysis of the SER performance utilizing the augmentation methods of this study with the existing augmentation methods in Table 9. We augmented the datasets without degrading the SER system performance. After DA, the updated data samples are 8400, 7200, 1920, 2140, and 44652 for TESS, RAVDESS, SAVEE, EMO-DB, and CREMA-D datasets, respectively.

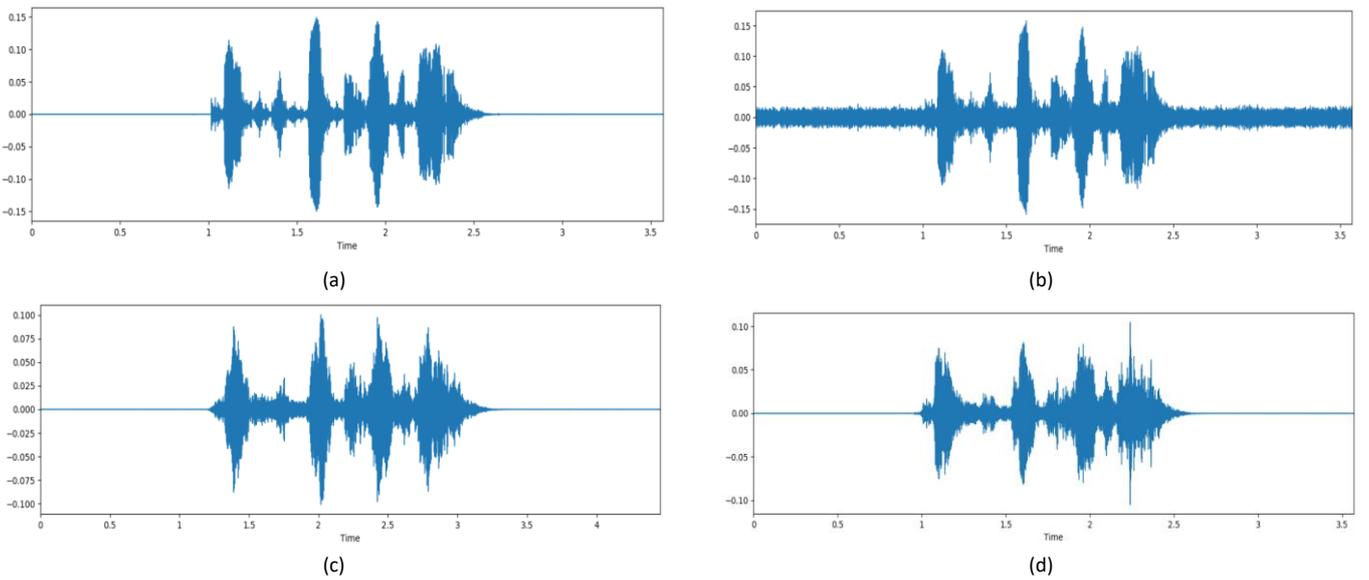

**Fig. 8.** A pictorial illustration of how various data augmentation techniques impacts speech utterances. Here, (a) is the original sound waveform, (b) is the AWGN injected waveform, (c) is the time-stretched waveform, and (d) is the waveform with shifted pitch.

*4.3. Feature extraction*

Extracting salient features from speech audio signals is one of the most important measures in SER-related activities. Precise extraction of crucial features improves the performance in terms of the SER accuracy of the model. Traditionally it is observed that low-level handcrafted features contain significant emotional cues about speech utterances and with proper feature engineering, work well with 1D CNN architecture (S. Zhang, Tao, Chuang, & Zhao, 2021). Properly configured 1D CNN architecture with a combination of LSTM, Bi-LSTM, GRU, and Bi-GRU architectures can perform effective feature engineering to acquire both local and global contextual cues from handcrafted speech features, and achieve excellent SER performance (J. Zhao et al., 2019). Specifically, this study uses five different spectral features: MFCC, LMS, ZCR, Chromagram, and RMS values of the speech audio files as the input for the proposed dilated 1D CNNs-FCNs, 1D CNNs-LSTM-FCNs, 1D CNNs-GRU-FCNs, and an ensemble of those three models. The brief details of the extracted features are given below.

**Mel-Frequency Cepstral Coefficients (MFCC):** Human-generated sounds are filtered through the vocal tract shape that includes tongue and teeth elements, which also is unique for each individual. The structure of these elements determines the voice of an individual. A precise measurement of the shape represents the phoneme being created. This shape is exhibited in the short-time power spectrum envelope, which is represented by MFCCs, and this feature is commonly used in SER research (Abdel-Hamid, 2020; Hajarolasvadi & Demirel, 2019; Z. T. Liu, Xie, et al., 2018; Nantasri et al., 2020). The MFCC feature extraction process is depicted in Fig. 9. It starts with the speech signal being converted into a short frame of 20-30ms window, and every 10ms, it is advanced, allowing the temporal features of individual speech signals to be traced. Then Discrete Fourier Transform (DFT) is performed on every windowed frame, and they are converted into magnitude spectrum using Eq. (12).

$$x_i(k) = \sum_{n=0}^{N-1} x_i(n)h(n)e^{\frac{-j2\pi kn}{N}} \qquad 0 \leq k \leq N-1 \qquad (12)$$

Here, $h(n)$ is the hamming window, $k$ which defines the DFT length, $x(n)$ represents the time-domain signal, $i$ defines the frame number, and $N$ defines the number of points used to calculate the DFT. After that, applying 26 filters in the previous signal the Mel-Scaled Filter-bank (MSFB) is calculated. MSFB is a measurement unit that is dependent on the frequency perception of the human ear. As a result, we have 26 numbers that describe the energy of each frame. The log energies are then calculated to obtain log filter-bank energies. The estimation of Mel from the physical frequency can be quantified through Eq. (13).

$$f_{Mel} = 2590 \log_{10}(1 + \frac{f}{700}) \qquad (13)$$

Here, $f$ denotes the physical frequency (Hz) and $f_{Mel}$ denotes the frequency perception of the human ear. Finally, Discrete Cosine Transform (DCT) is performed to get the MFCCs from the log filter-bank energies. For this study, 13-lower dimensions MFCCs were extracted from each audio file. Envelopes are sufficient to reflect the differences between phonemes, allowing us to recognize phonemes using MFCC. The sampling rate was set at 44.1 kHz, with DCT-2.

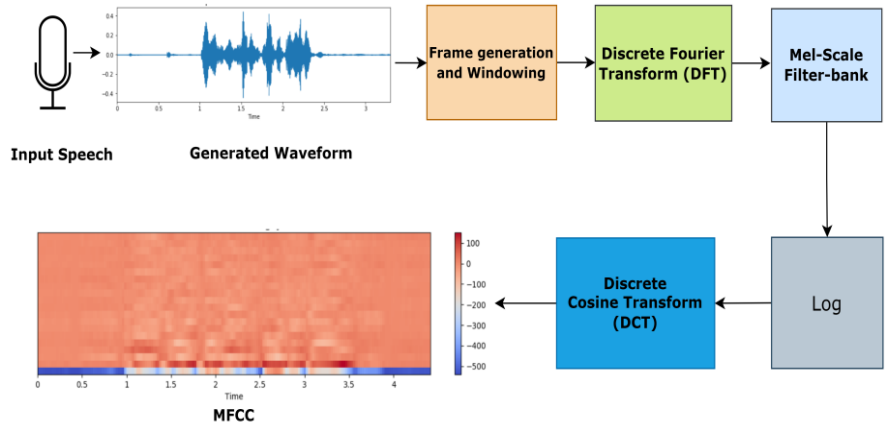

**Fig. 9.** A graphical illustration of the MFCC feature extraction process.

**Chromagram and Pitch:** The Chromagram is a time-frequency transformation of an acoustic signal into a briefly changing predecessor of the pitch and is used extensively in the SER task (Birajdar & Patil, 2020; Issa et al., 2020). It is related to the twelve diverse classes of the pitch. Applying Short-Time Fourier Transforms (STFT) to the waveform created from dataset audio files Chromagram features are collected. For this study, 12 Chromagram-bins were extracted from each audio file. The sound wave's frequencies determine the pitch feature in the SER task (Noroozi, Sapiński, Kamińska, & Anbarjafari, 2017). While the frequency is high, the pitch is considered high, and when the frequency is low, the pitch is considered as low too. In this study, the pitch factor was set at 0.6 and 0.7 during DA to create more samples for the training.

**Log-Mel Spectrogram (LMS):** The spectrogram portrays a signal's intensity in terms of the time-frequency domain. A spectrogram is generated by dividing a time-domain signal into equal-length segments. After that, each segment is subjected to the fast Fourier transform (FFT). The spectrogram is a plot of each segment's spectrum. It is a significant feature for any speech-related classification task and performs exceptionally with CNN (Hajarolasvadi & Demirel, 2019; Meng et al., 2019). For this study, 128 LMS features were extracted from each audio file. The use of multiple audio features rather than just one integrates several sound characteristics such as pitch, tone, harmony, etc., into a single training speech. This gives the SER models a more detailed interpretation of a speech sound sample, which improves their performance. A few of the randomly selected waveforms of the dataset's recordings and their corresponding spectrogram, MFCC, and Chromagram features are graphically represented in Fig. 10.

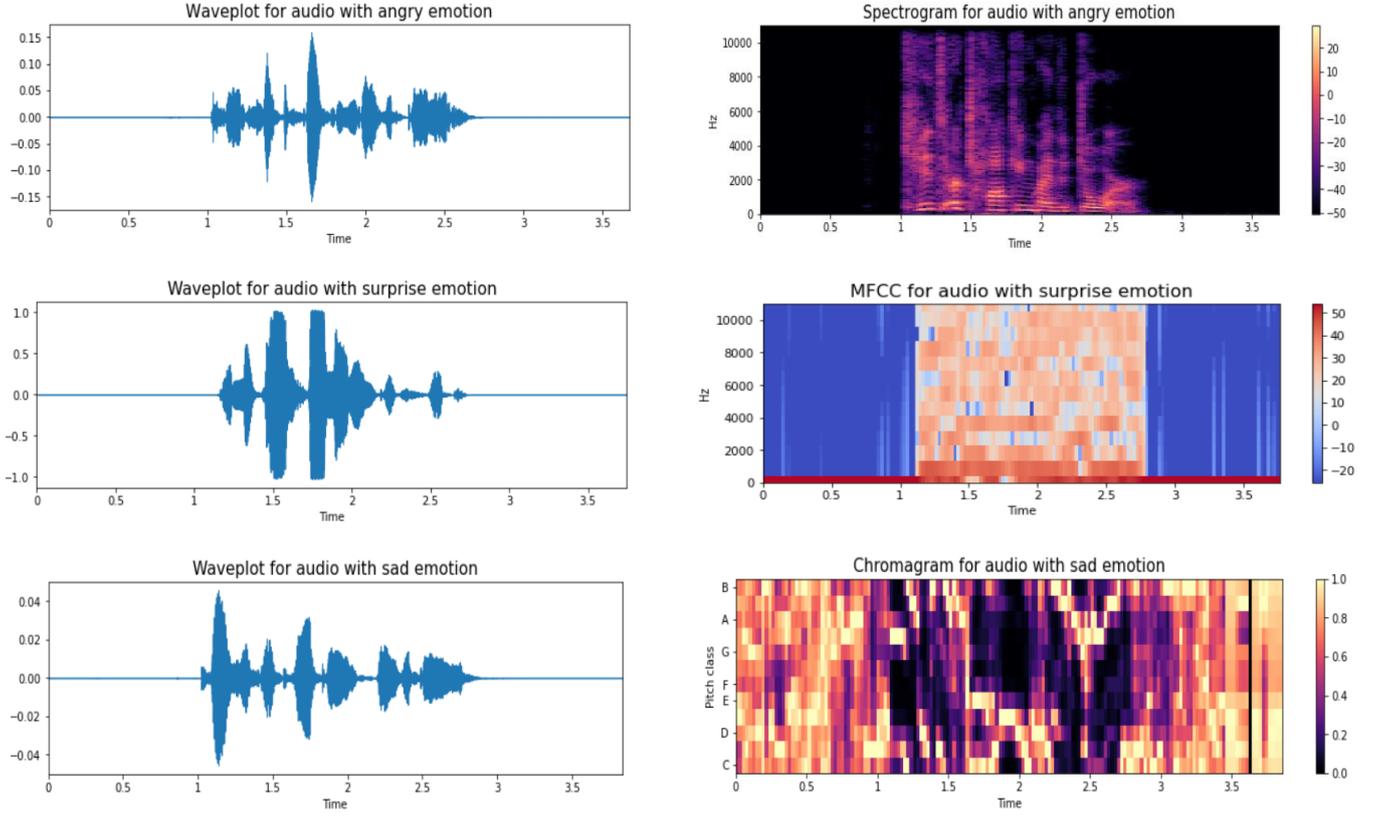

**Fig. 10**. Speech audio waveforms, and graphical representations of the spectrogram, MFCC, and Chromagram features of a few of the randomly selected categories of emotions from the experimented datasets.

**Zero Crossing Rate (ZCR):** ZCR is widely used for SER as well as music information collection-related tasks (Widiyanti & Endah, 2018). ZCR measures the number of times the amplitude of speech signals passes through zero value in a given period. ZCR is the best way to tell the difference between voiced and unvoiced expressions. There is no authoritative low-frequency fluctuation where there are frequent zero crosses. Mathematically, ZCR can be defined through Eq. (14), where $s$ denotes the signal of length $T$ and $1_{\mathbb{R}_{<0}}$ is an indicator function.

$$zcr = \frac{1}{T-1}\sum_{t=1}^{T-1} 1_{\mathbb{R}_{<0}}(s_t s_{t-1}) \quad (14)$$

**Root Mean Square (RMS) value:** It computes the RMS value for each frame from the speech audio samples. It performs an analysis of the overall amplitude of the signal, describing the average signal amplitude. RMS uses the magnitude of a signal as a measurement of signal strength, irrespective of the amplitude's positive or negative level. RMS and root mean square energy (RMSE) techniques are used by researchers (Mustaqeem & Kwon, 2020b; Yi, Mak, & Member, 2020) for speech audio that uses a signal's magnitude as a metric for signal power. For a given signal, $x = \{x_1, x_2, x_3,........x_n\}$, the RMS value, $x_{RMS}$ is calculated through Eq. (15).

$$x_{RMS} = \sqrt{\frac{x^2}{n}} = \sqrt{\frac{1}{n}(x_1^2 + x_2^2 + x_3^2 + ....... + x_n^2)} \quad (15)$$

This study's total number of extracted features is 13 MFCC, 12 Chromagram, 128 LMS, and two ZCR and RMS features, creating a feature vector of dimension 155 (128+13+12+1+1=155).

*4.4 Model training*

After getting the feature vector, the study performs data normalization by calculating the mean and standard deviation of the features. Data is divided into training data and testing parts with an 80:20 proportion. Those data are then turned into arrays and fed to the DL model as input. Since we deal with categorical data, each label is given a specific number dependent on alphabetical order. 20% of the data is used for model validation, and the remaining 80% of the data is used to train the models. Every individual model-A, B, and C is trained on both original datasets and augmented datasets. The whole process was carried out using the Keras framework for DL (Chollet, 2018). Grid-Search was applied to tune the hyper-parameter such as optimizer, batch size, learning rate, and weights to define the optimal values for all four models. According to Grid-Search's output, the batch size is set at 32, and weights were calculated to produce the highest WAA in the ensemble model-D. 'Adam' was selected as the optimizer, and we have chosen 'categorical cross entropy' as the loss function. Each individual model-A, B, and C was trained for 1000 epochs on Tesla P100-PCIE GPU. The prediction results of model-A, B, and C are then weighted and the ensemble model-D performs the final prediction based on those weighted models' predictive results. Assigning optimal weights to each individual model-A, B, and C provides the best results for the ensemble model-D.

## 5. Results Analysis

The number of correctly classified speech emotion labels (*True Positives-TP*), correctly classified instances that do not belong to the speech emotion label (*True Negatives-TN*), and instances that were either incorrectly classified to the speech emotion label (*False Positives-FP*) or were not classified as the speech emotion label (*False Negatives-FN*) will all be used to evaluate the correctness of a speech emotion classification task. A confusion matrix is made up of these four measurements (Sokolova & Lapalme, 2009). In this section, we present a detailed analysis of our experimental results. We show the training vs. validation accuracy curve and confusion matrix of every experiment performed in all five datasets for all three individual model-A, B, C, and ensemble model-D. Since the ensemble model-D performed best, we present the confusion matrix for this model both before and after performing DA. The confusion matrix for model-A, B, and C represents the model's performance after DA only.

### 5.1. Evaluation metrics

We analyze the performance of individual model-A, B, and C in terms of a weighted average (WA) score, since the datasets have an imbalanced distribution of classes. We also provide the precision, recall, and F1 score of the individual models in each dataset because of the class imbalance issues in those datasets. For the ensemble Model-D, we have utilized the WAA metric, which calculates accuracy by adjusting the weights of each model. For a distinct emotion label $L_i$, we define the evaluation by $TP_i$; $TN_i$; $FN_i$; $FP_i$; and Accuracy, Precision, and Recall are computed from the counts for $L_i$.

The accuracy metric presents the overall efficacy of the SER classifier (Sokolova & Lapalme, 2009).

$$\text{Accuracy} = \frac{TP+TN}{TP+FN+TN+FP} \tag{16}$$

The WAA metric computes the average accuracy by assigning weights to each individual model of the ensemble Model-D classifier (Z. Zhao et al., 2021).

$$\text{WAA} = \frac{\sum_{i=1}^{K}\left(\frac{TP_i+TN_i}{TP_i+FN_i+TN_i+FP_i}\right)}{K} \tag{17}$$

The precision metric calculates the number of $TP_i$ recognition that fall into the positive speech emotion labels ($TP_i + FP_i$) in a multiclass SER task.

$$\text{Precision} = \frac{\sum_{i=1}^{K} TP_i}{\sum_{i=1}^{K} TP_i + \sum_{i=1}^{K} FP_i} \tag{18}$$

The recall represents the proportion of correctly recognized positive speech emotion labels across all labels.

$$\text{Recall} = \frac{\sum_{i=1}^{K} TP_i}{\sum_{i=1}^{K} TP_i + \sum_{i=1}^{K} FN_i} \tag{19}$$

F1-score combines precision and recalls into a single metric that captures properties of both in a multiclass SER task.

$$\text{F1-score} = \frac{2*\sum_{i=1}^{K} \text{Precision}_i * \sum_{i=1}^{K} \text{Recall}_i}{\sum_{i=1}^{K} \text{Precision}_i + \sum_{i=1}^{K} \text{Recall}_i} \tag{20}$$

Macro-F1 calculates the F1-score for each class in the dataset and returns the average value without considering the percentage for each label without using any weights. All class is treated equally (Prasanth, Roshni Thanka, Bijolin Edwin, & Nagaraj, 2021).

$$\text{Macro-F1} = \frac{\frac{2*\sum_{i=1}^{K}\text{Precision}_i * \sum_{i=1}^{K}\text{Recall}_i}{\sum_{i=1}^{K}\text{Precision}_i + \sum_{i=1}^{K}\text{Recall}_i}}{K} \qquad (21)$$

Similarly, we can obtain the macro-recall and macro-precision scores by calculating the within-category values (Tan, 2005).

$$\text{Macro-recall} = \frac{\frac{\sum_{i=1}^{K}TP_i}{\sum_{i=1}^{K}TP_i + \sum_{i=1}^{K}FN_i}}{K} \qquad (22)$$

$$\text{Macro-precision} = \frac{\frac{\sum_{i=1}^{K}TP_i}{\sum_{i=1}^{K}TP_i + \sum_{i=1}^{K}FP_i}}{K} \qquad (23)$$

*5.2. Performance analysis*

All four models performed exceptionally well in each of the evaluated metrics. Table 2 presents the proposed individual model-A, B, and C's mean accuracy and ensemble model-D's weighted average accuracy in terms of SER performance with and without performing DA in each of the utilized datasets. At first, each model's performance is evaluated in the original dataset. All four models performed remarkably well in the original TESS dataset with over 96% mean weighted average accuracy. After performing DA, the performance improved further and achieved a WAA of 99.46% in the ensemble model-D. For the EMO-DB dataset, the SER performance of all four models was not up to the mark and drew the issue of overfitting, which is a common issue when deep models are trained on a relatively smaller size dataset. Another reason for this is the class distribution imbalance of the EMO-DB dataset and the low number of samples for the deep models to train efficiently. The same reason applies to the SAVEE dataset, with the WAA being poor for the original dataset with a very low number of samples for training the models. However, the performance of all the models significantly improved after performing DA and trained with an increased number of samples. The performance increased by around 32% from the non-augmented EMO-DB dataset and 22% from the non-augmented SAVEE dataset. In the EMO-DB dataset, the emotion "neutral" is often classified as "boredom," after rechecking the audio files, we saw that the spectral entropy properties of these two categories are pretty similar; that is why all the models are misclassifying these two types. The same reasoning goes for the SAVEE dataset for the emotions "happy" and "surprise." After rechecking, we found that some recordings with the "happy" labels are high pitched, almost the same as the emotion "surprise." However, the highest WAA achieved by the ensemble model-D in the EMO-DB and SAVEE datasets is 95.42% and 93.22%, respectively. When tested against the RAVDESS dataset, the models performed well, with the highest WAA achieved in the original dataset, and the augmented dataset is 89.19% and 95.62%, respectively. CREMA-D is the least explored dataset in SER-related studies. It is challenging to work with because many actors and actresses from different races uttered different sentences in the dataset. The human accuracy of this dataset's audio-only part is around 40.9%. All four models exceeded that number, with the ensemble achieving around 68.14% WAA in the original dataset and 90.47% WAA in the augmented dataset. Tested against the original CREMA-D dataset, the "neutral" and "disgust" emotions are often misclassified as "sad" because both have a lower pitch and amplitude in the signal waveform; with similar spectral properties. When we trained all three individual models with the increased augmented data, all the discussed issues were resolved significantly. The accuracy curve for training versus validation after DA is also presented for all four models in Fig. 11-15. The confusion matrix for the ensemble model-D is achieved by adjusting the proper weights of model-A, B, and C after performing a grid-search technique. In Fig. 16, we provide the confusion matrix of the ensemble model-D to evaluate the performance before and after data augmentation in each of the utilized datasets.

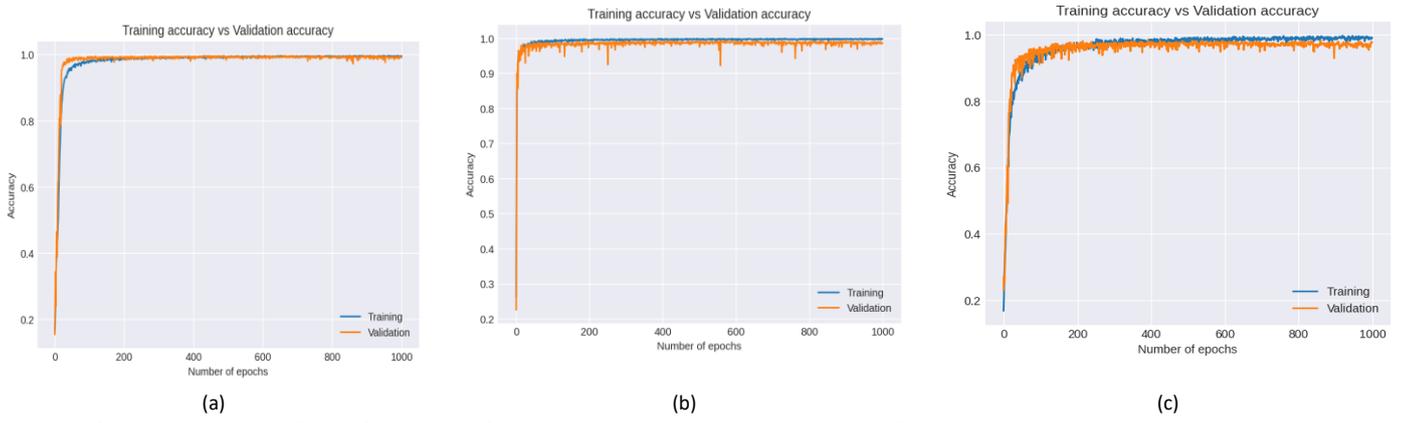

**Fig. 11.** Performance evaluation (after performing DA) of the proposed models in the TESS dataset. (a), (b), and (c) show the training vs. validation accuracy curve for model-A, model-B, and model-C, respectively, trained for 1000 epochs.

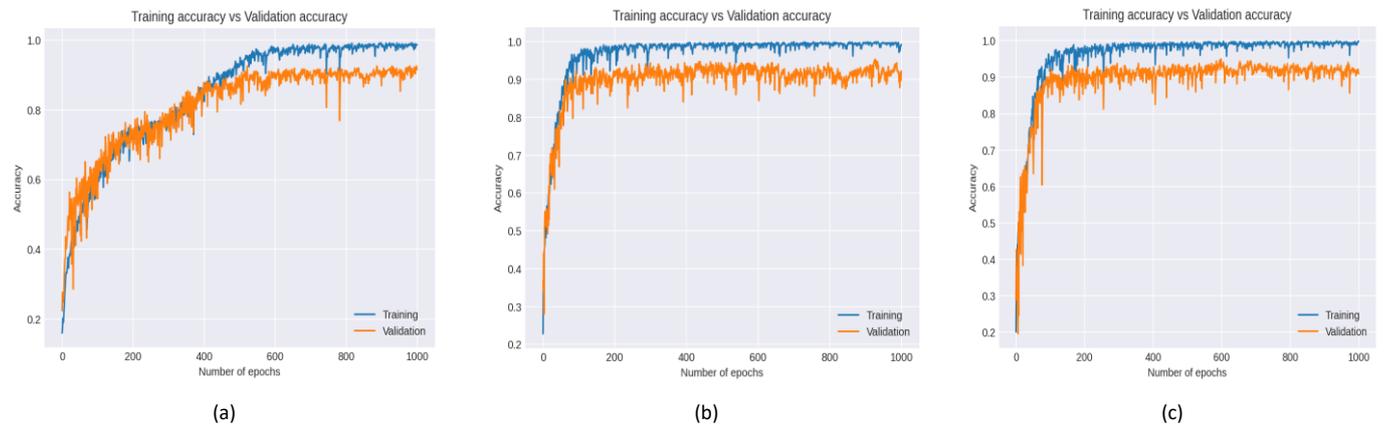

**Fig. 12**. Performance evaluation (after performing DA) of the proposed models in the EMO-DB dataset. (a), (b), and (c) show the training vs. validation accuracy curve for model-A, model-B, and model-C, respectively, trained for 1000 epochs.

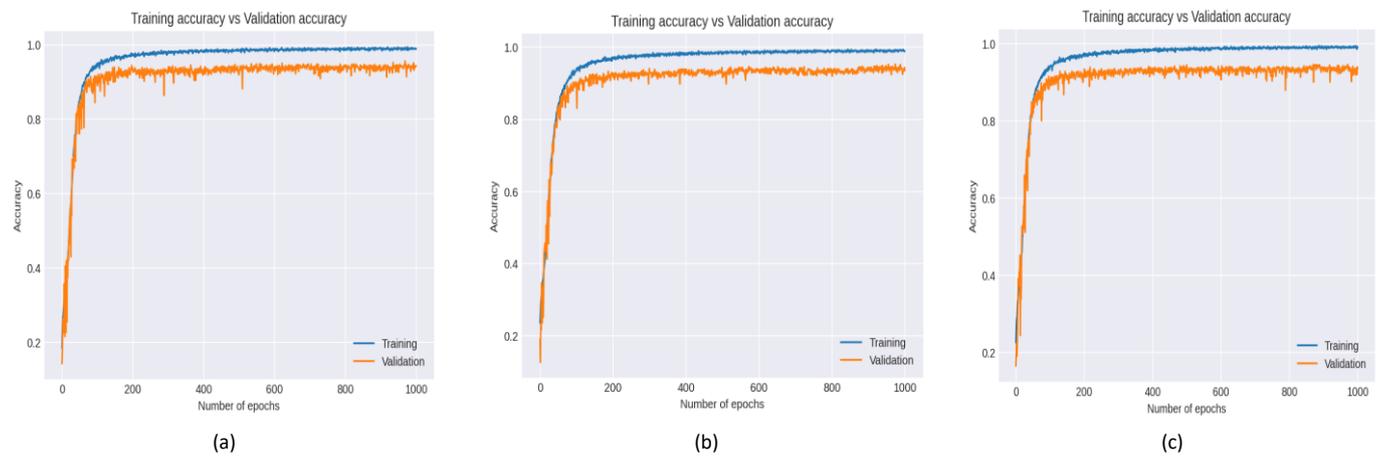

**Fig. 13**. Performance evaluation (after performing DA) of the proposed models in the RAVDESS dataset. (a), (b), and (c) show the training vs. validation accuracy curve for model-A, model-B, and model-C, respectively, trained for 1000 epochs.

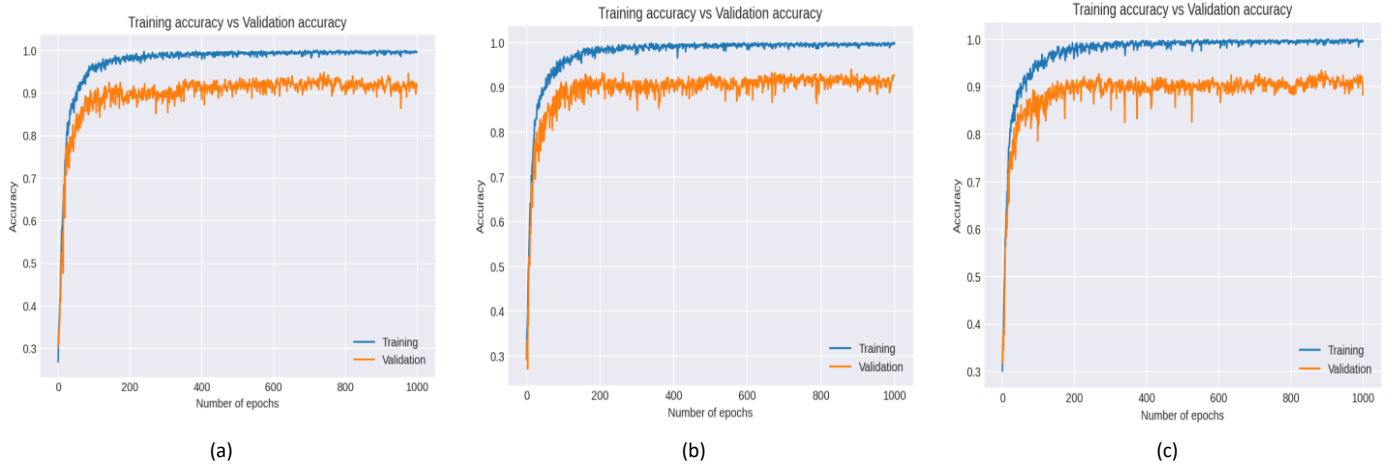

**Fig. 14.** Performance evaluation (after performing DA) of the proposed models in the SAVEE dataset. (a), (b), and (c) show the training vs. validation accuracy curve for model-A, model-B, and model-C, respectively, trained for 1000 epochs.

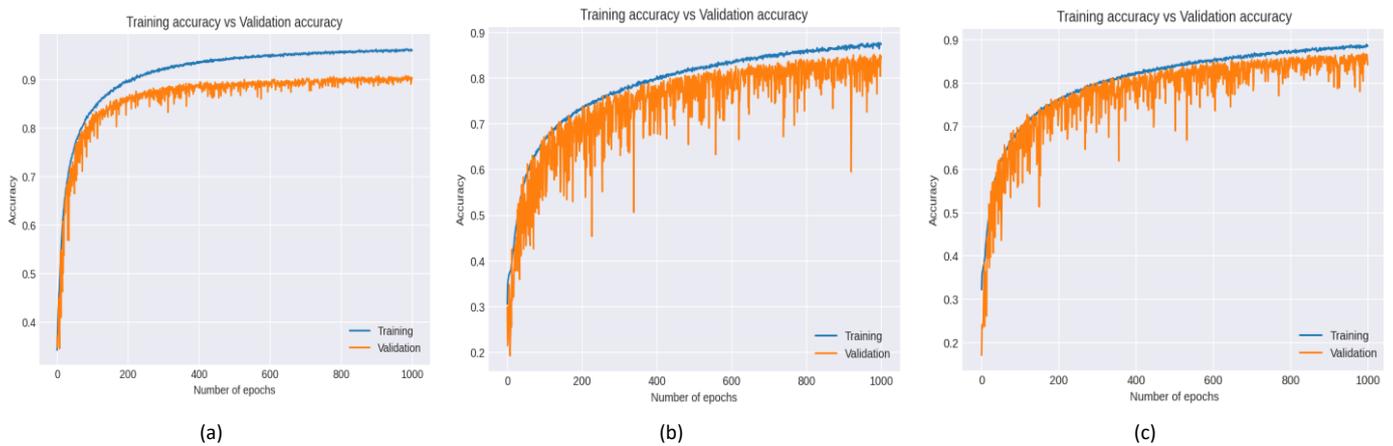

**Fig. 15.** Performance evaluation (after performing DA) of the proposed models in the CREMA-D dataset. (a), (b), and (c) shows the training vs. validation accuracy curve for model-A, model-B, and model-C, respectively, trained for 1000 epochs.

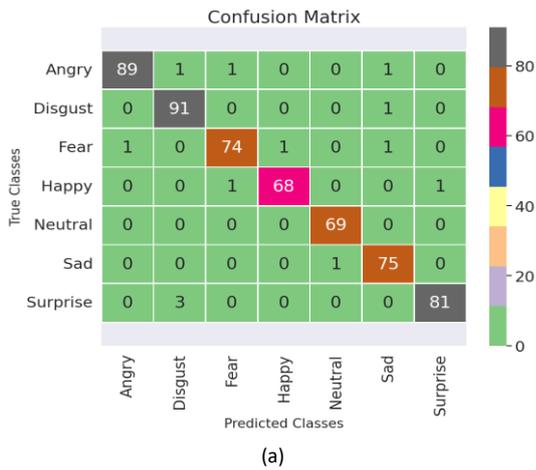
(a)

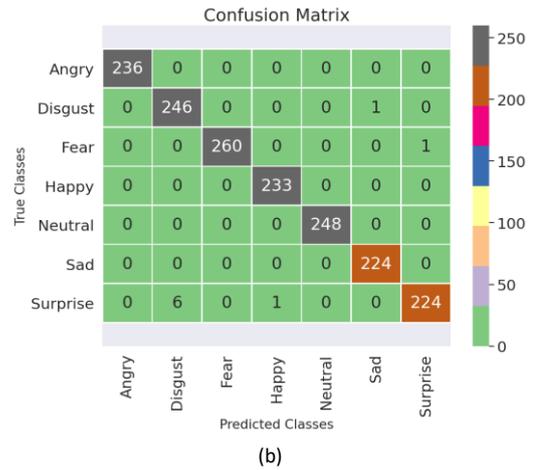
(b)

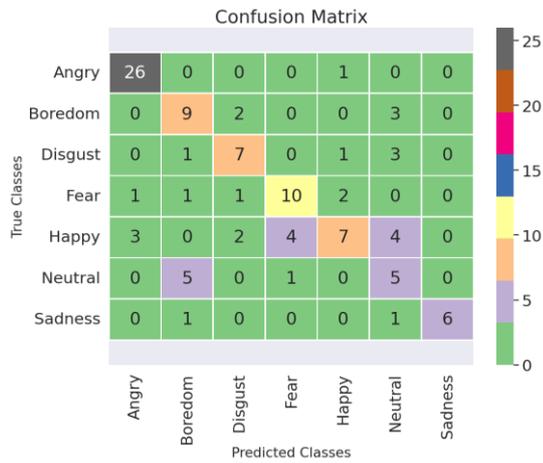

(c)

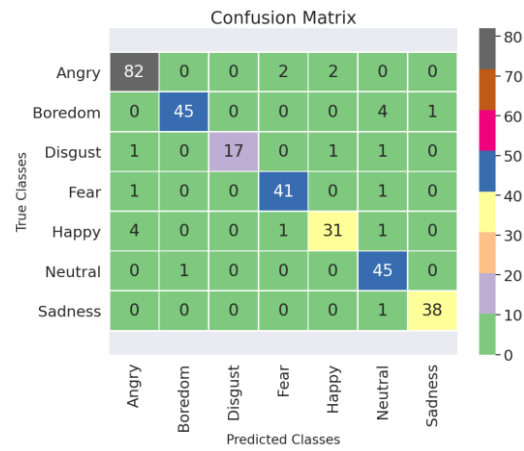

(d)

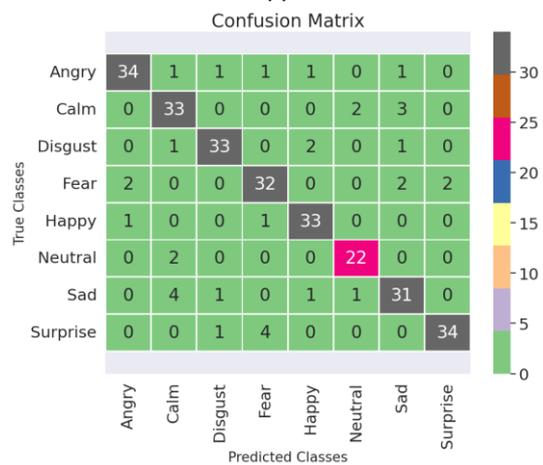

(e)

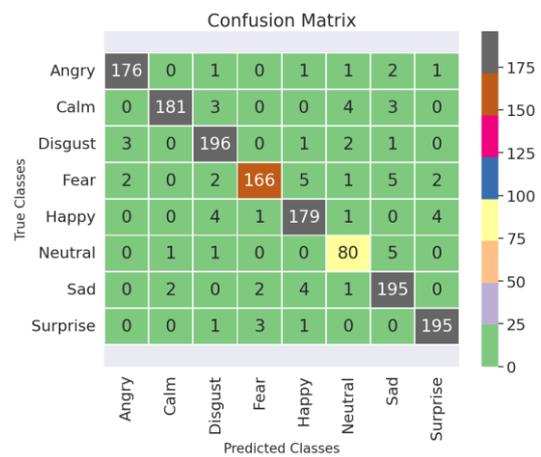

(f)

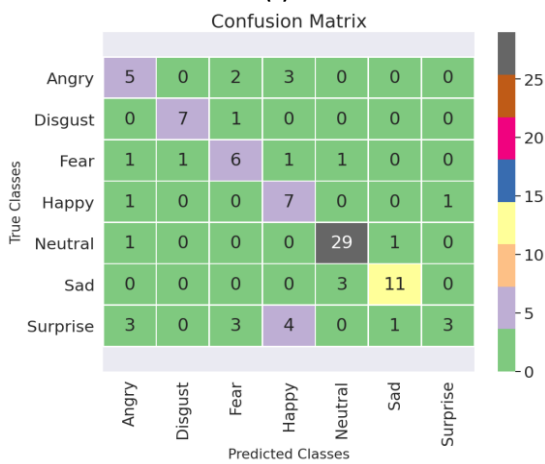

(g)

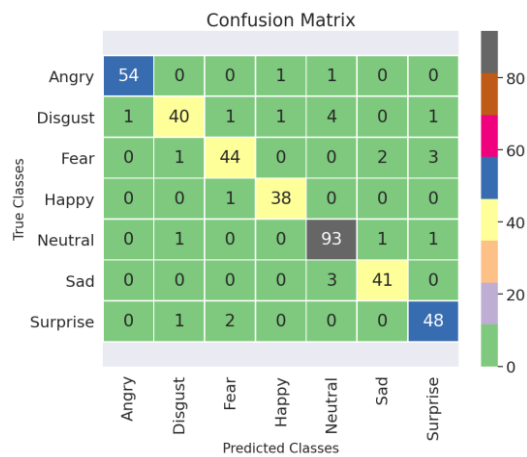

(h)

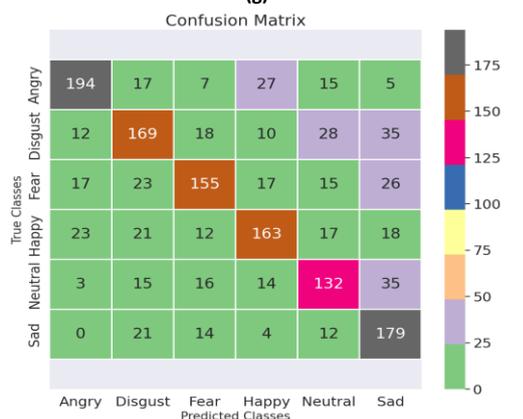

(i)

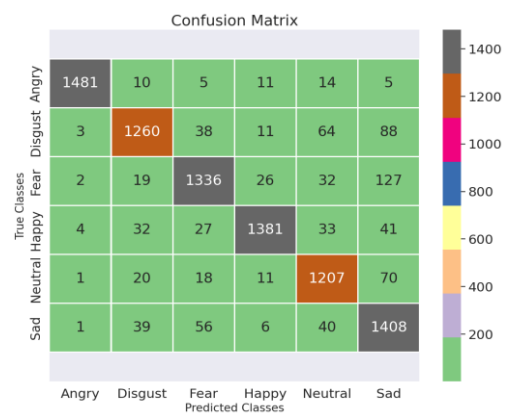

(j)

**Fig. 16.** Performance evaluation of the proposed ensemble model-D both before and after performing data augmentation (DA) in the utilized datasets. Here, (a) and (b) present the confusion matrix for the TESS dataset before and after performing DA, respectively. (c) and (d) present the confusion matrix for the EMO-DB dataset before and after performing DA, respectively. (e) and (f) present the confusion matrix for the RAVDESS dataset before and after performing DA, respectively. (g) and (h) present the confusion matrix for the SAVEE dataset before and after performing DA, respectively. (i) and (j) present the confusion matrix for the CREMA-D dataset before and after performing DA, respectively.

**Table 2**

Comparison of all four proposed models based on the SER performance on TESS, EMO-DB, RAVDESS, SAVEE, and CREMA-D datasets.

| Datasets | Model Name | Mean Accuracy (Model-A, B, C) / Weighted Average Accuracy (WAA) (Ensemble Model-D) | |
|---|---|---|---|
| | | Without data augmentation (%) | With data augmentation (%) |
| TESS | Model-A | 96.78 | 99.05 |
| | Model-B | 96.00 | 98.49 |
| | Model-C | 95.68 | 98.10 |
| | Weighted Ensemble model-D | 98.00 | 99.46 |
| EMO-DB | Model-A | 65.88 | 92.00 |
| | Model-B | 64.32 | 92.21 |
| | Model-C | 65.18 | 91.53 |
| | Weighted Ensemble model-D | 67.74 | 95.42 |
| RAVDESS | Model-A | 86.11 | 94.38 |
| | Model-B | 88.54 | 93.61 |
| | Model-C | 86.77 | 94.00 |
| | Weighted Ensemble model-D | 89.19 | 95.62 |
| SAVEE | Model-A | 68.00 | 92.00 |
| | Model-B | 65.87 | 93.00 |
| | Model-C | 68.14 | 88.28 |
| | Weighted Ensemble model-D | 71.00 | 93.22 |
| CREMA-D | Model-A | 66.60 | 90.22 |
| | Model-B | 66.00 | 84.27 |
| | Model-C | 65.88 | 84.39 |
| | Weighted Ensemble model-D | 68.14 | 90.47 |

Table 3 presents the class-wise SER performance of each individual model-A, B, and C after performing DA in terms of precision, recall, F1-score, mean accuracy, average macro-precision, macro-recall, and macro-F1 for the TESS, EMO-DB, RAVDESS, SAVEE, and CREMA-D datasets. Fig. 17 illustrates the overall performance of model-A, B, and C in terms of their average macro precision, macro recall, mean accuracy, and macro F1-score values. Macro precision and macro recall values are significantly high for the TESS and RAVDES datasets. One reason is that all the models converge well in those datasets because of their superior balanced class distribution compared to EMO-DB and SAVEE datasets. Baseline model-A achieves the highest macro precision score in all the utilized datasets. In terms of macro recall, model-B achieves the highest results apart from the CREMA-D dataset only. Model-C's overall performance is lower than the other models, especially in the EMO-DB and SAVEE datasets. Model-B achieves the highest mean accuracy and macro F1-score value in all the utilized datasets, except for the CREMA-D dataset. However, for the CREMA-D dataset, model-A performs significantly better than other individual models. It is observed that the performance of model-C is poor compared to the other two individual models in both mean accuracy and macro F1-score metrics.

**Table 3**

Class-wise SER performance of individual model-A, B, and C on the utilized datasets. The best results per dataset are highlighted in **bold** font.

| TESS dataset | | | | | | | | | | | | |
|---|---|---|---|---|---|---|---|---|---|---|---|---|
| | Model-A | | | | Model-B | | | | Model-C | | | |
| Category | Precision (%) | Recall (%) | F1 (%) | Mean Accuracy (%) | Precision (%) | Recall (%) | F1 (%) | Mean Accuracy (%) | Precision (%) | Recall (%) | F1 (%) | Mean Accuracy (%) |
| Angry | 100 | 100 | 100 | | 100 | 100 | 100 | | 100 | 100 | 100 | |
| Disgust | 97 | 99 | 98 | | 98 | 100 | 99 | | 96 | 99 | 98 | |
| Fear | 100 | 98 | 100 | | 98 | 100 | 94 | | 94 | 100 | 100 | |
| Happy | 98 | 100 | 99 | 99 | 99 | 100 | 99 | **99.40** | 99 | 100 | 99 | 99.10 |
| Neutral | 100 | 98 | 100 | | 100 | 100 | 100 | | 100 | 100 | 100 | |
| Sad | 100 | 100 | 100 | | 96 | 100 | 99 | | 97 | 100 | 100 | |
| Surprise | 100 | 95 | 97 | | 100 | 97 | 98 | | 99 | 95 | 97 | |
| *Macro Average* | **99** | 98 | 98 | | 98 | **99** | 98 | | 97 | **99** | **99** | |
| **EMO-DB dataset** | | | | | | | | | | | | |
| Angry | 92 | 97 | 94 | | 96 | 95 | 95 | | 95 | 94 | 95 | |
| Boredom | 96 | 86 | 91 | | 92 | 90 | 90 | | 95 | 76 | 84 | |
| Disgust | 100 | 85 | 92 | | 86 | 88 | 88 | | 89 | 85 | 87 | |
| Fear | 95 | 88 | 92 | 92.26 | 95 | 95 | 95 | **92.38** | 86 | 98 | 91 | 91.66 |
| Happy | 86 | 84 | 85 | | 83 | 92 | 87 | | 94 | 86 | 90 | |
| Neutral | 81 | 96 | 88 | | 89 | 91 | 90 | | 86 | 93 | 90 | |
| Sadness | 97 | 97 | 97 | | 97 | 92 | 95 | | 89 | 100 | 94 | |
| *Macro Average* | **93** | 90 | **91** | | 91 | **92** | 91 | | 91 | 90 | 90 | |
| **RAVDESS dataset** | | | | | | | | | | | | |
| Angry | 97 | 96 | 96 | | 96 | 94 | 94 | | 96 | 97 | 96 | |
| Calm | 98 | 94 | 96 | | 96 | 96 | 96 | | 96 | 96 | 96 | |
| Disgust | 92 | 97 | 94 | | 98 | 89 | 93 | | 90 | 90 | 90 | |
| Fear | 96 | 90 | 93 | **94.38** | 98 | 92 | 95 | 94 | 97 | 92 | 94 | 93.86 |
| Happy | 92 | 95 | 94 | | 89 | 99 | 93 | | 97 | 94 | 95 | |
| Neutral | 88 | 90 | 90 | | 83 | 93 | 86 | | 83 | 92 | 87 | |
| Sad | 93 | 93 | 93 | | 92 | 92 | 92 | | 93 | 93 | 93 | |
| Surprise | 95 | 97 | 97 | | 93 | 99 | 96 | | 93 | 95 | 94 | |
| *Macro Average* | 93 | 94 | **94** | | 93 | **95** | 93 | | 93 | 94 | 93 | |
| **SAVEE dataset** | | | | | | | | | | | | |
| Angry | 96 | 93 | 95 | | 98 | 96 | 97 | | 96 | 95 | 95 | |
| Disgust | 100 | 85 | 92 | | 91 | 88 | 89 | | 92 | 69 | 79 | |
| Fear | 91 | 82 | 86 | | 90 | 88 | 89 | | 78 | 78 | 78 | |
| Happy | 91 | 100 | 100 | 92 | 95 | 97 | 96 | **92.70** | 92 | 87 | 89 | 88.28 |
| Neutral | 92 | 97 | 97 | | 92 | 96 | 94 | | 84 | 98 | 90 | |
| Sad | 81 | 89 | 89 | | 93 | 89 | 91 | | 89 | 91 | 90 | |
| Surprise | 92 | 94 | 94 | | 90 | 92 | 91 | | 94 | 90 | 92 | |
| *Macro Average* | 92 | 91 | 92 | | **93** | **92** | **93** | | 89 | 87 | 88 | |
| **CREMA-D dataset** | | | | | | | | | | | | |
| Angry | 99 | 97 | 98 | | 98 | 95 | 96 | | 97 | 95 | 96 | |
| Disgust | 93 | 85 | 88 | | 83 | 78 | 81 | | 84 | 81 | 82 | |
| Fear | 89 | 87 | 88 | 90.22 | 86 | 77 | 81 | 84.27 | 86 | 76 | 81 | 84.39 |
| Happy | 95 | 91 | 93 | | 93 | 87 | 90 | | 92 | 85 | 89 | |
| Neutral | 88 | 91 | 89 | | 84 | 79 | 81 | | 75 | 86 | 81 | |
| Sad | 83 | 90 | 86 | | 68 | 90 | 77 | | 72 | 87 | 78 | |
| *Macro Average* | **91** | **90** | **90** | | 85 | 84 | 84 | | 84 | 85 | 84 | |

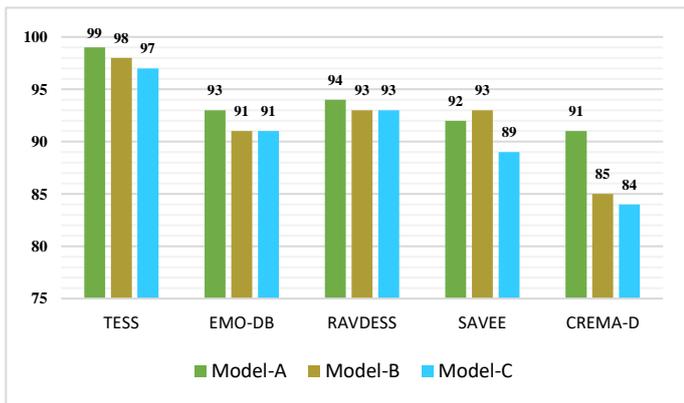

(a)

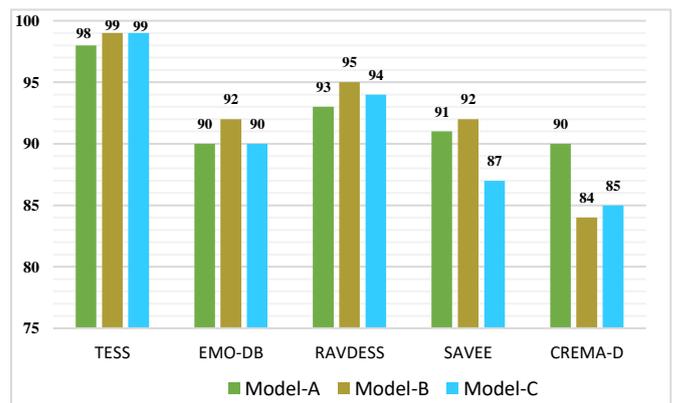

(b)

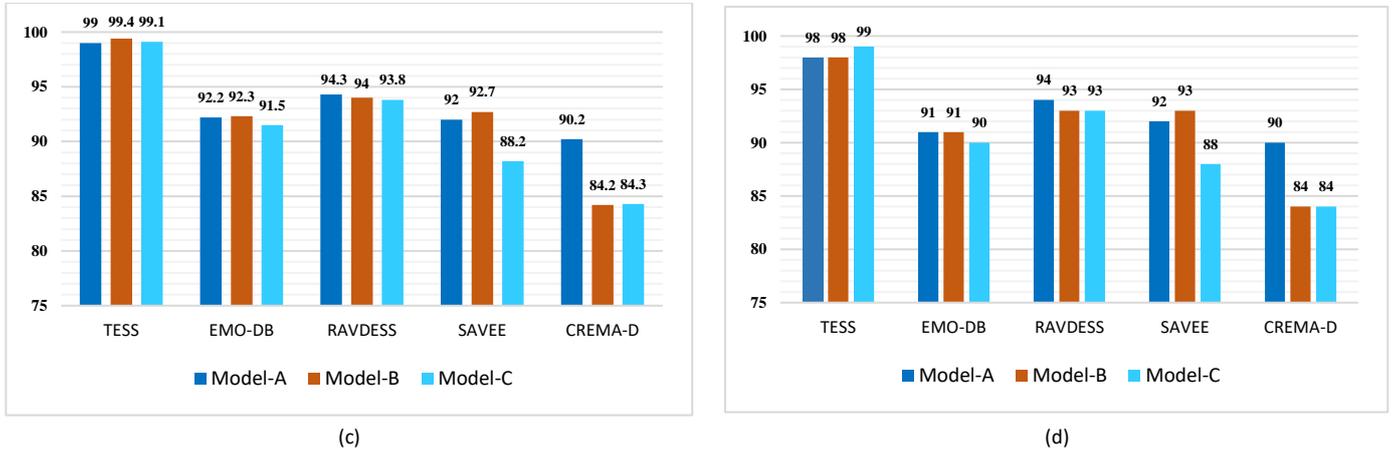

**Fig. 17.** Graphical bar plot depictions of the individual models - A, B, and C in terms of (a) macro precision (b) macro recall (c) mean accuracy, and (d) macro F1-score. All the values are represented in terms of percentage.

*5.3. Comparative analysis with other methods*

There has already been extensive research in the field of SER. However, comparing performance was tough because only a few performed DA in those datasets for SER (Padi, Manocha, & Sriram, 2020; Tiwari et al., 2020; Yi et al., 2020) using GAN or multi-window method, and by adding generative noise. We have identified that augmenting data is necessary for the utilized datasets because the sizes of these datasets are significantly lesser for the proper training of a DL-based model. Besides that, only a few bits of literature utilized the TESS and CREMA-D datasets for this task. Some of the existing articles (Demircan & Kahramanli, 2018; Hajarolasvadi & Demirel, 2019; S. Li et al., 2021) use only a subset of those datasets; some perform feature extraction from the audio, text, and video samples of those datasets (Ristea, Dutu, & Radoi, 2019; Yoon et al., 2018). Our scope in this research is only audio samples. Some (Chen, He, Yang, & Zhang, 2018; Kim, Englebienne, Truong, & Evers, 2017; Mustaqeem & Kwon, 2021c) evaluate their framework's performance using a different metric from ours, like unweighted average accuracy, recall (K. Feng & Chaspari, 2020; Meng et al., 2019); Some choose a questionable training and testing split ratio of 90:10 (Bhavan et al., 2019); therefore, we only compare with those articles that match our criterion.

In Table 4-8, we present the performance comparison of our work with previous work for TESS, EMO-DB, RAVDESS, SAVEE, and CREMA-D datasets, respectively. Table 9 compares our work with those articles adopting different data augmentation methods to increase the SER accuracy using the utilized datasets of this study. The comparison shows that this study's data augmentation approach uses a lesser feature dimension than other methods, providing improved results.

In addition to the better SER performance achieved by the proposed DL-based model−A, B, C and weighted ensemble model-D in all the utilized datasets (see Table 2-9), the training time complexity and size of our proposed models are relatively lightweight and occupy less memory compared to other reported SOTA SER architectures. The training time of model-A, B, and C are 3685 s, 3929 s, and 3798 s on the TESS dataset, 23429 s, 25170 s, and 25475 s on the CREMA-D dataset, 972 s, 989 s, and 981 s on the EMO-DB dataset, 3038 s, 3241 s, and 3108 s on the RAVDESS dataset, and 919 s, 931 s, and 922 s on the SAVEE dataset, respectively. Though it should be noted that, training time varies based on the utilized GPU's and allocated memory. In terms of model size, all three proposed models-A, B, and C are less memory consuming compared to reported SER benchmarks such as CB-SER (Mustaqeem et al., 2020), ADRNN (Meng et al., 2019), ATT-Net (Mustaqeem & Kwon, 2021b), ACRNN (Chen et al., 2018), QCNN (Muppidi & Radfar, 2021), and DSCNN (Mustaqeem & Kwon, 2020a). Size of the models- A, B, and C are 23.5 MB, 38 MB, and 34.5 MB on the TESS dataset, 23.4 MB, 37.1 MB, and 33.8 MB on the CREMA-D dataset, 20.5 MB, 22.1 MB, and 21.6 MB on the RAVDESS dataset, 19.8 MB, 21.7 MB, and 21.1 MB on the EMO-DB dataset, and 18.5 MB, 20 MB, and 19.3 MB on the SAVEE dataset, respectively. Due to the increased model complexity used in the individual models, we trade off the ensemble model-D's evaluation time for robust SER performance, which can only be completed after training all three separate model-A, B, and C. For all datasets, each of the proposed individual models exhibits better generalization during the experimental assessment, ensuring higher recognition accuracy with minimal computation cost. The lightweight property makes all the models suitable for real-time applications for human-computer interaction. Among all four models, ensemble model-D performed best in terms of SER accuracy in all the datasets. The individual model's excellent performance in detecting emotion from speech across all five datasets and adjusting the proper weights for each model for the ensemble prediction contributes mainly to the improved recognition rate of weighted ensemble model-D.

**Table 4**
Performance comparison of this work with recent literature in the TESS dataset.

| Reference | Methodology | Features | Accuracy |
|---|---|---|---|
| (Mekruksavanich, Jitpattanakul, & Hnoohom, 2020) | DCNN | MFCC | 55.71% |
| (R. Chatterjee et al., 2021) | 1D-CNN | MSFB-Cepstral Coefficients | 95.79% |
| (Praseetha & Vadivel, 2018) | DNN, RNN, GRU | MFCC, LMS | 95.82% |
| (Aggarwal et al., 2022) | DNN, VGG-16 | 2D LMS | 97.15% |
| (Venkataramanan & Rajamohan, 2019) | 2D CNN | LMS | 62.00% |
| This work | 1D CNNs-FCNs | MFCC, LMS, ZCR, Chromagram, and RMS value | 99.05% |
| | 1D CNNs-LSTM-FCNs | | 98.40% |
| | 1D CNNs-GRU-FCNs | | 98.10% |
| | Ensemble Model-D | | 99.46% |

**Table 5**
Performance comparison of this work with recent literature in the EMO-DB dataset.

| Reference | Methodology | Features | Accuracy |
|---|---|---|---|
| (Issa et al., 2020) | 1D CNN | MFCC, LMS, Chromagram, Spectral contrast, Tonnetz | 86.10% |
| (Tiwari et al., 2020) | DNN | ZCR, RMS energy, MFCC, and statistical features | 82.73% |
| (Yadav & Vishwakarma, 2020) | 1D CNN, Bi-LSTM | Acoustic features | 94.00% |
| (J. Zhao et al., 2019) | 1D-2D DCNN, LSTM | Spectral features | 95.33% |
| (Anvarjon et al., 2020) | 2D CNN | LMS | 92.02% |
| (Mustaqeem et al., 2020) | Bi-LSTM | LMS | 85.57% |
| (Mustaqeem & Kwon, 2021b) | CNN, Channel Attention | LMS | 93.00% |
| (D. Li et al., 2021) | Bi-LSTM | MFCC, Spectral centroid, roll-off, flux, and spread, ZCR, RMS, Chromagram, Pitch, entropy | 85.95% |
| (Ancilin & Milton, 2021) | SVM | Mel Frequency Magnitude Coefficient | 81.50% |
| (Farooq et al., 2020) | DCNN, SVM, MLP | LMS | 95.10% |
| (Nantasri et al., 2020) | ANN | MFCCs, Delta, Delta-Deltas | 87.80% |
| (Yi et al., 2020) | DNN, SVM, GAN, Autoencoder | MFCC, ZCR, RMS | 84.49% |
| This work | 1D CNNs-FCNs | MFCC, LMS, ZCR, Chromagram, and RMS value | 92.26% |
| | 1D CNNs-LSTM-FCNs | | 92.38% |
| | 1D CNNs-GRU-FCNs | | 91.66% |
| | Ensemble Model-D | | 95.42% |

**Table 6**
Performance comparison of this work with recent literature in the RAVDESS dataset (- Not mentioned).

| Reference | Methodology | Features | Accuracy |
|---|---|---|---|
| (Issa et al., 2020) | 1D CNN | MFCC, LMS, Chromagram, Spectral contrast, Tonnetz | 71.61% |
| (Yadav & Vishwakarma, 2020) | 1D CNN, Bi-LSTM | Acoustic features | 73.00% |
| (Mekruksavanich et al., 2020) | DCNN | MFCC | 75.83% |
| (Farooq et al., 2020) | DCNN, SVM, MLP | LMS | 81.30% |
| (Padi et al., 2020) | CNN | MFCC, Chromagram, and Time-domain features | 88.00% |
| (Nantasri et al., 2020) | ANN | MFCCs, Delta, Delta-Deltas | 82.30% |
| (Mustaqeem et al., 2020) | Bi-LSTM | LMS | 77.02% |
| (Mustaqeem & Kwon, 2020b) | 1D CNN | - | 80.00% |
| (Mustaqeem & Kwon, 2021b) | CNN, Channel Attention | LMS | 80.00% |
| (Ancilin & Milton, 2021) | SVM | Mel frequency magnitude coefficient | 64.31% |
| (Aggarwal et al., 2022) | DNN | MFCC, Chromagram, LMS, Spectral centroid and roll-off | 73.95% |
| This Work | 1D CNNs-FCNs | MFCC, LMS, ZCR, Chromagram, and RMS value | 94.38% |
| | 1D CNNs-LSTM-FCNs | | 94.00% |

|  | 1D CNNs-GRU-FCNs |  | 93.86% |
|  | Ensemble Model-D |  | 95.62% |

**Table 7**
Performance comparison of this work with recent literature in the SAVEE dataset.

| Reference | Methodology | Features | Accuracy |
|---|---|---|---|
| (Hajarolasvadi & Demirel, 2019) | 3D CNN | LMS | 81.05% |
| (Farooq et al., 2020) | DCNN, SVM, MLP | LMS | 82.10% |
| (Padi et al., 2020) | CNN | MFCC, Chromagram, and Time-domain features | 70.00% |
| (Ancilin & Milton, 2021) | SVM | Mel frequency magnitude coefficient | 75.63% |
| (Mekruksavanich et al., 2020) | DCNN | MFCC | 65.83% |
| (Z. T. Liu, Xie, et al., 2018) | GA, PCA, LLD | MFCC | 76.40% |
| This work | 1D CNNs-FCNs | MFCC, LMS, ZCR, Chromagram, and RMS value | 92.00% |
| This work | 1D CNNs-LSTM-FCNs |  | 92.70% |
| This work | 1D CNNs-GRU-FCNs |  | 88.28% |
| This work | Ensemble Model-D |  | 93.22% |

**Table 8**
Performance comparison of this work with recent literature in the CREMA-D dataset (- Not mentioned).

| Reference | Methodology | Features | Accuracy |
|---|---|---|---|
| (Mekruksavanich et al., 2020) | DCNN | MFCC | 65.77% |
| (Singh et al., 2020) | SVM | MFCC, ZCR, RMSE | 58.22% |
| (Scheidwasser-clow, Kegler, Beckmann, Cernak, & Epfl, 2022) | Convolutional Transformer | - | 76.90% |
| (Mocanu & Tapu, 2021) | CNN | Acoustic features | 64.85% |
| (Huang, Tao, Liu, & Lian, 2020) | LSTM, Vector of Locally Aggregated Descriptors | MFCC, Low level descriptors | 63.50% |
| This work | 1D CNNs-FCNs | MFCC, LMS, ZCR, Chromagram, and RMS value | 90.22% |
| This work | 1D CNNs-LSTM-FCNs |  | 84.27% |
| This work | 1D CNNs-GRU-FCNs |  | 84.39% |
| This work | Ensemble Model-D |  | 90.47% |

**Table 9**
Performance comparison of this work with recent literature adopting different data augmentation techniques in the utilized datasets (- Not mentioned).

| Reference | Methodology | Datasets | Mean Accuracy/WAA | Features | Feature dimension | Data augmentation method |
|---|---|---|---|---|---|---|
| (Tiwari et al., 2020) | DNN | EMO-DB | 76.77% | ZCR, RMS energy, MFCC, and statistical features | 6552 | Generative noise model |
| (Yi et al., 2020) | DNN, SVM, GAN, Auto encoder | EMO-DB | 84.49% | MFCC, ZCR, RMS | 4368 | Adversarial data augmentation network |
| (S. Zhang et al., 2018) | Deep CNN | EMO-DB | 87.31% | 2D LMS | - | Increased overlap length of speech signals |
| (T. Feng, Hashemi, Annavaram, & Narayanan, 2022) | CNN, Adversarial learning | CREMA-D | 69.80% | 2D LMS | - | Addition of AWGN |
| (Praseetha & Joby, 2021) | GRU | TESS | 93.00% | Filter-bank Energies | - | Tempo and speed perturbation |
| (Padi et al., 2020) | CNN | SAVEE | 70.00% | MFCC, Chromagram, and Time-domain features | 34 | Multi-Window based method |
| (Padi et al., 2020) | CNN | RAVDESS | 88.00% | MFCC, Chromagram, and Time-domain features | 34 | Multi-Window based method |
|  | CNN, LSTM | RAVDESS | 92.60% | MFCC, ZCR, RMS |  |  |

| Reference | Model | Dataset | Accuracy | Features | # Features | Data Augmentation |
|---|---|---|---|---|---|---|
| (Jothimani, S and Premalatha, 2022) | | CREMA-D | 89.90% | | | Noise Removal, White Noise Injection, and Pitch Tuning |
| | | SAVEE | 84.90% | | | |
| | | TESS | 99.60% | | | |
| (Lalitha, Gupta, Zakariah, & Alotaibi, 2020) | MLP, RF | EMO-DB | 87.30% | MFCC, inverted MFCC, extended MFCC, extended IMFCC, LPC, Mel, and Bark filter bank-derived features | - | Synthetic Minority Over-sampling Technique (SMOTE) |
| | | SAVEE | 75.20% | | | |
| This work | 1D CNNs-FCNs | EMO-DB | 92.26% | MFCC, LMS, ZCR, Chromagram, and RMS value | 155 | Injecting AWGN, stretching the speech audio files, and modification of the pitch of the sound |
| | 1D CNNs-LSTM-FCNs | | 92.38% | | | |
| | 1D CNNs-GRU-FCNs | | 91.66% | | | |
| | Ensemble Model-D | | 95.42% | | | |
| | 1D CNNs-FCNs | SAVEE | 92.00% | | | |
| | 1D CNNs-LSTM-FCNs | | 92.70% | | | |
| | 1D CNNs-GRU-FCNs | | 88.28% | | | |
| | Ensemble Model-D | | 93.22% | | | |
| | 1D CNNs-FCNs | RAVDESS | 94.38% | | | |
| | 1D CNNs-LSTM-FCNs | | 94.00% | | | |
| | 1D CNNs-GRU-FCNs | | 93.86% | | | |
| | Ensemble Model-D | | 95.62% | | | |
| | 1D CNNs-FCNs | CREMA-D | 90.22% | | | |
| | 1D CNNs-LSTM-FCNs | | 84.27% | | | |
| | 1D CNNs-GRU-FCNs | | 84.39% | | | |
| | Ensemble Model-D | | 90.47% | | | |

## 6. Conclusion and Future Works

Inadequate data could prohibit any DNN-based model from achieving its maximum ability, which is a significant challenge in the DL-based SER task. The lack of data samples often leads a deep and complex model to suffer from overfitting issue. This paper presents a comprehensive study of different DL-based SER systems utilizing five different datasets, covering two languages: English and German. We have handcrafted five types of LLD features from each audio file. We have designed multiple LFABs inside the baseline model-A to learn local hidden features of the speech signals. An additional GFAB is added to both model-B and model-C that extracts long-term global contextual dependencies and correlations from the learned features of LFABs. The effectiveness of a weighted ensemble setting of three new DL-based models is assessed on five standard benchmark SER datasets. With data augmentation, the result of the proposed weighted ensemble model-D is significant, achieving a SOTA WAA of 99.46%, 95.42%, 95.62%, 93.22%, and 90.47% for the TESS, EMO-DB, RAVDESS, SAVEE, and CREMA-D datasets respectively.

Although there have been steady advancements in methods, features, and obtained accuracy in SER, many limitations are yet to be addressed for an effective and industry-grade SER scheme. The majority of the datasets are acted, scripted, and only cover a few discrete statements and expressions throughout the corpus. There can be major differences between working with real and acted data. Moreover, in most cases, the sample size in those datasets is insufficient to adequately train a DL-based model. Experiments conducted to create those datasets are simulated and semi-natural. They are not noisy and are far away from the natural environment in a real-world scenario. This brings questions about the ability of a developed system which are built using those datasets to detect the correct emotion in a real-world noisy scenario. The detection of emotions in a dialogue between multiple actors from continuous speech audio is an area that needs further research. A possible extension of this research is to develop a multi-label problem where an utterance in any conversation often contains multiple emotion types. The fusion of information from multi-modal CNN architectures that capture different optimal acoustic features from speech signals needs more addressing and further investigation. Most of the utilized acoustic features contain information about magnitude and phase. However, the traditional SER system mostly focuses on the magnitude information only. The exploration of effective phase-based features is another point of research direction. Even though this study performs exceptionally well in SER across five datasets, we believe that further analysis on this subject is necessary. In the future, we hope to reduce the training time needed for the individual models to make an ensemble prediction by focusing more on the optimal feature selection method and integrating different attention mechanisms to get more optimal cues for the SER task.

*CRediT authorship contribution statement*

**Md. Rayhan Ahmed:** Conceptualization, Methodology, Formal analysis, Validation, Investigation, Resources, Data curation, Writing – original draft and revision. **Salekul Islam:** Formal analysis, Validation, Writing – original draft. **A.K.M. Muzahidul Islam:** Formal analysis, Validation, Writing – original draft. **Swakkhar Shatabda:** Formal analysis, Validation, Writing – original draft.

**Declaration of Competing Interest**

The authors declare that they have no known competing financial interests or personal relationships that could have appeared to influence the work reported in this paper.